\documentclass[aps,reprint,superscriptaddress]{revtex4-1}
\usepackage{amssymb}
\usepackage{graphicx}
\usepackage{color}
\usepackage{amsmath}

\newcommand{\bfk}{{\bf k}}
\newcommand{\bfv}{{\bf v}}

\newcommand{\mueff}{\mu_{\mathrm{eff}}}
\newcommand{\gammaeff}{\gamma_{\mathrm{eff}}}

\begin{document}


\title{Analytic expressions for electron-ion temperature equilibration rates from the Lenard-Balescu equation}


\author{Christian R. Scullard}
\affiliation{Lawrence Livermore National Laboratory, Livermore CA 94550, USA}
\author{Susana Serna}
\affiliation{Departament de Matematiques, Universitat Aut\`onoma de Barcelona, 08193 Bellaterra-Barcelona, Spain}
\author{Lorin X. Benedict}
\author{C. Leland Ellison}
\author{Frank R. Graziani}
\affiliation{Lawrence Livermore National Laboratory, Livermore CA 94550, USA}
\email[]{scullard1@llnl.gov}


\date{\today}

\begin{abstract}
In this work, we elucidate the mathematical structure of the integral that arises when computing the electron-ion temperature equilibration time for a homogeneous weakly-coupled plasma from the Lenard-Balescu equation. With some minor approximations, we derive an exact formula, requiring no input Coulomb logarithm, for the equilibration rate that is valid for moderate electron-ion temperature ratios and arbitrary electron degeneracy. For large temperature ratios, we derive the necessary correction to account for the coupled-mode effect, which can be evaluated very efficiently using ordinary Gaussian quadrature.
\end{abstract}


\maketitle

\section{Introduction}
Computing the equilibration time of a two-temperature electron-ion plasma is a fundamental problem in plasma physics. Over the decades, theories have been developed that include physics beyond what is captured in simple Landau-Spitzer formulas \cite{LS,LS2}, such as fermion statistics and collective oscillations. For a weakly-coupled plasma, the quantum Lenard-Balescu equation \cite{LB,qLB} is believed \cite{Gericke05,Daligault,Benedict} to be a very good approximation. In the form given in Refs.~\cite{Gericke05,Vorberger,Daligault,Benedict}, the equilibration rate is
\begin{eqnarray}
& &\frac{d T_i}{dt}= -\frac{\hbar}{3 \pi^3 n_i} \int_0^{\infty} dk k^2 \int_0^{\infty} d\omega \omega \left|\frac{v_{ei}(k)}{\epsilon(k,\omega)} \right|^2 \cr
&\times& \left[N \left(\frac{\hbar \omega}{2 k_B T_i}\right)-N \left(\frac{\hbar \omega}{2 k_B T_e}\right) \right] \mathrm{Im} \chi_e^{(0)}(k,\omega) \mathrm{Im} \chi_i^{(0)}(k,\omega), \cr \label{eq:dTdti}
& &
\end{eqnarray}
where $N(x)=\coth(x)$, $\chi_j^{(0)}(k,\omega)$ is the free-particle response function of species $j$, $v_{ei}(k)$ is the Fourier transform of the Coulomb potential,
\begin{equation}
 v_{ei}(k)=-\frac{4 \pi Z e^2}{k^2},
\end{equation}
$Z$ is the ionic charge, assumed fixed throughout our discussion, and $\epsilon(k,\omega)$ is the dielectric function in the random phase approximation,
\begin{equation}
 \epsilon(k,\omega)=1-\frac{4 \pi e^2}{k^2}[\chi_e^{(0)}(k,\omega)+Z^2 \chi_i^{(0)}(k,\omega)] . \label{eq:eps_definition}
\end{equation}
Equation (\ref{eq:dTdti}) can be derived in many different ways; linear response and fluctuation-dissipation arguments \cite{Daligault}, Keldysh Green's functions \cite{Dharma-wardana,Chapman}, or simply starting from the quantum Lenard-Balescu collision operator \cite{Gericke05,Vorberger}.

It is not our aim here to shed any new light on the physics of this problem. Rather, we devote this work to developing techniques for evaluating (\ref{eq:dTdti}) efficiently enough that it can be quickly carried out as part of a larger computation, such as a radiation-hydrodynamic simulation of a fusion-burning plasma \cite{ICF}. In the process we will show that these integrals are not nearly as difficult as generally assumed, especially with the help of some fairly minor approximations, and indeed we give an analytic formula, equation (\ref{eq:main}), that is accurate over a wide range of conditions. Where (\ref{eq:main}) is not necessarily a good approximation is deep in the so-called coupled mode regime \cite{Vorberger}. This occurs when the electron and ion temperatures are separated sufficiently, and/or $Z$ is sufficiently large, that an ion acoustic oscillation impacts the rate. Mathematically, this occurs in the part of the integrand where
\begin{equation}
 \mathrm{Re}\ \epsilon(k,\omega)=0
\end{equation}
and $\mathrm{Im}\epsilon(k,\omega)$ is small. This will manifest as a sharp peak in the integrand for values of $k$ and $\omega$ where the latter conditions are satisfied. As we will show, the approximation leading to (\ref{eq:main}) begins to break down when
\begin{equation}
 T_i < 0.28 Z T_{\mathrm{eff}} \label{eq:CMcond}
\end{equation}
where $T_{\mathrm{eff}}$ is an effective electron temperature that we will derive carefully later. Suffice it to say for the moment that for strongly degenerate electrons \cite{Vorberger}
\begin{equation}
 T_{\mathrm{eff}} \rightarrow \frac{2}{3}E_F ,
\end{equation}
where $E_F$ is the Fermi energy (see (\ref{eq:Theta}) below), and
\begin{equation}
 T_{\mathrm{eff}} \rightarrow T_e
\end{equation}
in the classical limit. Vorberger and Gericke \cite{Vorberger} give a condition similar to (\ref{eq:CMcond}) although they estimate the constant to be more like $0.27$. This difference is unimportant, and one might even question the value of stating the coupled mode condition (\ref{eq:CMcond}) with such precision; we only wish to point out that the particular coupled mode effect of interest to us only begins causing mathematical issues when (\ref{eq:CMcond}) is satisfied. In practice, our main concern regarding coupled modes is computing (\ref{eq:dTdti}) when $T_i \ll Z T_{\mathrm{eff}}$.

We should point out here that recent work \cite{Benedict17} has called into question whether coupled modes can ever really impact the temperature equilibration rate in a physical plasma system. Specifically, the effect was not seen in classical molecular dynamics simulations set up at conditions where it was expected to be important. Instead, the observed rate was closer to that obtained from the so-called Fermi golden rule approximation \cite{Dharma-wardana}, in which the dielectric function is factored into electron and ion pieces,
\begin{equation}
 |\epsilon(k,\omega)|^2=|\epsilon_e(k,\omega)|^2 |\epsilon_i(k,\omega)|^2 . \label{eq:FGR}
\end{equation}
This approximation facilitates analytic computation using a sum rule but decouples the modes. It was argued in Ref. \cite{Benedict17} that strong ion-ion coupling may modify the predictions of the standard Lenard-Balescu equation, in which the random phase approximation is adopted for the plasma screening \cite{LB,qLB}. Although some of us were involved in that work, we will take no position on it here. Instead, we present a method for efficiently correcting (\ref{eq:main}) to account for the coupled mode effect when it might occur (and as predicted using a strict application of the Lenard-Balescu equation \cite{Dharma-wardana,Vorberger,Chapman}), and leave the debate about the necessity of such a correction for another place. This will enable researchers to perform sensitivity studies in which the potential effects of coupled modes can be assessed in applications, following the spirit of Ref.~\cite{Garbett}.

Our main approximation will be the neglect of quantum diffraction {\it in the dielectric function only}. We have scant evidence that its inclusion has an appreciable impact on the calculation of equilibration rates and, somewhat surprisingly perhaps, it simplifies the problem enormously. Another important mathematical issue revolves around the coefficient $\alpha \equiv m_e T_i/(m_i T_e)$. The large mass disparity between electrons and ions means that $\alpha$ is generally small for physically relevant situations, and when $T_e \ll T_i$ it only becomes smaller. In section \ref{sec:alphazero}, we show that when we set $\alpha=0$ and neglect diffraction in the dielectric function, it is possible to find an exact expression for (\ref{eq:dTdti}). The strategy is to find the Laurent expansion of the dielectric function and to perform the integral in the complex frequency plane. These details are relegated to the appendices, with the main text used to report the results and to call attention to particularly interesting features of the derivation. The exact formula is given in powers of $1/\Lambda$, where $\Lambda$ is a parameter we will derive that is typically large in weak coupling. It will also prove to be the argument of the ubiquitous Coulomb logarithm \cite{LS,LS2}, $\log \Lambda$, although we stress that this quantity will arise naturally from the evaluation of convergent integrals and will not be put in ``by hand'' as it is in the Landau-Spitzer formula.

Where coupled modes are important, setting $\alpha=0$ in the dielectric function prevents one from fully capturing the effect. In this regime, the method of section \ref{sec:alphazero} does not work and in section \ref{sec:coupledmode} we lay out the modified strategy for handling this situation. Here, rather than evaluating the $\omega$-integral exactly, we first perform the $k$-integral to order $1/\Lambda^3$, leaving us with a one-dimensional integral over $\omega$ that has no difficult peaks but can no longer be solved exactly. It can, however, be handled with ordinary Gaussian quadrature. This allows us to isolate the place it is necessary to retain a non-zero $\alpha$ and so to derive a correction to be added, if needed, to the $\alpha=0$ formula.

In section \ref{sec:numerical}, we compare numerical evaluations of (\ref{eq:dTdti}) to our exact solution to demonstrate its wide range of validity.

\section{Integrand} \label{sec:integral}
Here, we define the various functions used in (\ref{eq:dTdti}) and the approximations that facilitate our calculations. We work in dimensionless variables with the help of the following definitions,
\begin{eqnarray}
& &x^2 \equiv \frac{m_i \beta_i \omega^2}{2 k^2},\ 
 y^2 \equiv \frac{\hbar^2 \beta_i k^2}{8 m_e},\ 
 \alpha \equiv \frac{m_e \beta_e}{m_i \beta_i} \cr 
& & \gamma \equiv \frac{\beta_e}{\beta_i}=\frac{T_i}{T_e},\ 
 \eta \equiv \frac{\lambda_Q}{\lambda_D},\ \rho \equiv 2 \sqrt{m_e/m_i} (\gamma-1) \label{eq:defs}
\end{eqnarray}
and
\begin{equation}
 \lambda_Q^2 \equiv \frac{\hbar^2 \beta_i}{8 m_e},\ 
 \lambda_D^2 \equiv \frac{1}{4 \pi e^2 Z^2 n_i \beta_i} \  .
\end{equation}
Note that, as already pointed out, the parameter $\alpha$ is always small unless $T_i \gg T_e$ by a factor comparable to the mass ratio, an extreme situation that will not concern us here.
\subsection{Response functions}
The free-particle response function is given by the integral
\begin{eqnarray}
\chi^{(0)}_q(k,\omega) = \lim_{\eta \rightarrow 0^+} \int d^3 {\bfv} \frac{f(\bfv)-f(\bfv+\hbar \bfk/m)}{\hbar \omega - \hbar \bfv \cdot \bfk - \frac{\hbar^2 k^2}{2 m}+i \eta} . \label{eq:chi}
\end{eqnarray}
For the electrons, we use the Fermi-Dirac distribution,
\begin{equation}
 f_e(v)=\frac{2 m_e^3}{h^3} \frac{1}{\exp \left(\beta_e m_e v^2/2+\mu \right)+1} \ , \label{eq:FD}
\end{equation}
which we write in terms of the classical momentum $m_e v$ when doing the response function integrals. The dimensionless chemical potential, $\mu$, is determined by particle number conservation,
\begin{equation}
 \int_0^\infty \frac{x^{1/2} dx}{1+e^{x+\mu}}=\frac{2}{3}\Theta^{-\frac{3}{2}} \label{eq:mueq}
\end{equation}
where $\Theta$ is the degeneracy parameter
\begin{equation}
 \Theta \equiv \frac{1}{\beta_e E_F}=\frac{2 m_e}{\beta_e \hbar^2 (3 \pi^2 n_e)^{2/3}} \ . \label{eq:Theta}
\end{equation}
Note that our $\mu$ is the negative of the usual definition; a fit is given in Appendix \ref{sec:Fermi_functions}.

The imaginary part of the electron response function is well-known (e.g., \cite{Ichimaru}) and is given by
\begin{eqnarray}
 \mathrm{Im} \chi^{(0)}_e(x,y)&=&-\frac{m_e^2}{2 \pi \hbar^4 \beta_e k} \left[\ln \left(1+e^{-\mu-(\sqrt{\alpha}x-\sqrt{\gamma}y)^2} \right) \right.\cr
 &-&\left. \ln \left(1+e^{-\mu-(\sqrt{\alpha}x+\sqrt{\gamma}y)^2} \right) \right] \ . \label{eq:Imchi_e}
\end{eqnarray}
This form includes both quantum diffraction and electron degeneracy. Expanding (\ref{eq:Imchi_e}) to lowest order in the small parameter $\alpha$, we have the alternative form that we will use in (\ref{eq:dTdti}) in the numerator, outside the dielectric function,
\begin{equation}
  \mathrm{Im} \chi_e(x,y)\approx -\frac{m_e^{3/2}\sqrt{\alpha}x}{\sqrt{2}\pi \sqrt{\beta_e}\hbar^3} \frac{1}{1+e^{\gamma y^2+\mu}} \ . \label{eq:Imchie_sm_alpha}
\end{equation}
As for the ions, we use (\ref{eq:chi}) but with the Maxwell distribution
\begin{equation}
 f_i(v)=n_i \left(\frac{m_i \beta_i}{2 \pi} \right)^{3/2} \exp \left(-\beta_i m_i v^2\right)
\end{equation}
which leads to
\begin{equation}
 \mathrm{Im} \chi_i(x,y)=-\frac{n_i \beta_i \sqrt{\pi}}{2}\sqrt{\frac{m_i}{m_e}}\frac{1}{y}e^{-y^2-x^2} \sinh \left(2 \sqrt{\frac{m_e}{m_i}}xy \right).
\end{equation}
Once again, this will be used outside the dielectric function in (\ref{eq:dTdti}).

\subsection{Dielectric function}
Inside the dielectric function, as promised, we drop quantum diffraction. That is, we take the limit $\hbar \rightarrow 0$ where it explicitly appears in (\ref{eq:chi}) but retain it in the Fermi-Dirac distribution. We need both the real and imaginary parts of the response functions, and for the ions we have the well-known expressions
\begin{eqnarray}
 \mathrm{Re} \chi_i(x) &=& - n_i \beta_i \left[1 -  2 x F(x) \right] \label{eq:Rechi_i} \\
 \mathrm{Im} \chi_i(x) &=& - n_i \beta_i \sqrt{\pi} x e^{-x^2} \ , \label{eq:Imchi_i}
\end{eqnarray}
where $F(x)$ is the Dawson function \cite{Dawson},
\begin{equation}
 F(x) \equiv e^{-x^2} \int_0^x e^{t^2}dt = x e^{-x^2}\int_0^1 e^{s^2 x^2} ds \ . \label{eq:Dawson}
\end{equation}
In Appendix \ref{sec:response_functions}, we derive the electron response function,
\begin{eqnarray}
& &\mathrm{Re}\chi_e(x) =  \cr
 & & \ \ \ \ \  - n_e \beta_{\mathrm{eff}} \left[1 - \frac{e^{-\mu}}{I_{-\frac{1}{2}}(-\mu)}2 \sqrt{\alpha} x F(\sqrt{\alpha} x;\mu) \right] \label{eq:Rechi_e_eff}\\
& &\mathrm{Im} \chi_e(x) = - n_e \beta_{\mathrm{eff}} \sqrt{\pi} \frac{1}{I_{-\frac{1}{2}}(-\mu)} \frac{\sqrt{\alpha}x}{1+e^{\mu} e^{\alpha x^2}} \label{eq:Imchi_e_eff} \ ,
\end{eqnarray}
where $I_{-\frac{1}{2}}(-\mu)$ is a Fermi-Dirac integral, defined in equation (\ref{eq:Iminhalf}), and $F(x;\mu)$ is a generalization of the Dawson function for degenerate electrons. As we will see in section \ref{sec:coupledmode}, we can set $\alpha=0$ in the real part, even in the coupled mode regime, so we fortunately never need to evaluate $F(x;\mu)$. The effective electron temperature, $\beta_{\mathrm{eff}}$, is defined to make (\ref{eq:Rechi_e_eff}) and (\ref{eq:Imchi_e_eff}) look as much like their classical counterparts, (\ref{eq:Rechi_i}) and (\ref{eq:Imchi_i}), as possible. Comparing (\ref{eq:Rechi_e_static}) to (\ref{eq:Rechi_e_nodiff}) (with $\alpha=0$) gives
\begin{equation}
 n_e \beta_{\mathrm{eff}} \equiv \frac{4 \sqrt{2} \pi^{3/2} m_e^{3/2}I_{-\frac{1}{2}}(-\mu)}{h^3 \sqrt{\beta_e}} \ , \label{eq:betaeff}
\end{equation}
and we also define an effective temperature ratio,
\begin{equation}
 \gamma_{\mathrm{eff}} \equiv \frac{\beta_{\mathrm{eff}}}{\beta_i} .
\end{equation}
The effective temperature has the limits
\begin{equation}
\begin{array}{lcl}
 \beta_{\mathrm{eff}} \rightarrow \beta_e & \mbox{when}& T_e \gg T_F \\
 \beta_{\mathrm{eff}} \rightarrow 1/(k_B T_F) & \mbox{as}& T_e \rightarrow 0 \ .
\end{array}
\end{equation}
Note that an effective electron temperature is often used in Coulomb logarithms in the form
\begin{equation}
 T_{\mathrm{eff}}=\left[T_e^p+\left( \frac{2}{3}E_F \right)^p \right]^{1/p} \label{eq:Teff_approx}
\end{equation}
precisely to capture these two limits. Generally, $p=2$ is used but in Ref. \cite{Stanton} it was suggested that $p=9/5$ produces slightly better results for some calculations. We find as well that $p=9/5$ provides a very accurate approximation to (\ref{eq:betaeff}), with maximum error around 2\%, although using $p=2$ is not very much worse. On the other hand, our formula can be easily evaluated with the help of Dandrea, Ashcroft and Carlsson's \cite{Dandrea} very accurate Pad\'e approximant, given in Appendix \ref{sec:Fermi_functions}. Putting the results of this section together, the dielectric function is
\begin{equation}
 \epsilon(x,y)=1+\frac{\eta^2}{y^2}w(x) \label{eq:dielectric}
\end{equation}
where
\begin{equation}
 w(x)=w_r(x)+i w_i(x)
\end{equation}
and
\begin{eqnarray}
 w_r(x)&=&1+\frac{\gamma_{\mathrm{eff}}}{Z} - 2xF(x) \label{eq:wr} \\
 w_i(x)&=&\sqrt{\pi} \left (\frac{\gamma_{\mathrm{eff}}}{Z}\frac{1}{I_{-\frac{1}{2}}(-\mu)} \frac{\sqrt{\alpha}x}{1+e^{\mu} e^{\alpha x^2}} +x e^{-x^2} \right). \label{eq:wi}
\end{eqnarray}
Here we can clearly see the benefit of dropping quantum diffraction. Normally, $w(x)$ would be a function of both $x$ and $y$, as is obvious from a glance at (\ref{eq:Imchi_e}), but instead we have (\ref{eq:dielectric}). This clean separation between the variables $x$ and $y$ is a key component of our (otherwise) exact solution.

\subsection{Final form of integral}
Now, we put the response functions (\ref{eq:Imchie_sm_alpha}) and (\ref{eq:Imchi_i}) into the integral (\ref{eq:dTdti}). Making use of the following identity,
\begin{eqnarray}
& & \left[N \left(\frac{\hbar \omega}{2 k_B T_i}\right)-N \left(\frac{\hbar \omega}{2 k_B T_e}\right) \right]\sinh \left(\frac{\hbar \omega \beta_i}{2}\right) \cr
&=& \frac{\sinh\left[2(\gamma-1)\sqrt{m_e/m_i}xy\right]}{\sinh\left[2 \sqrt{m_e/m_i} \gamma xy\right]} \cr
&\approx& \frac{\sinh\left[2(\gamma-1)\sqrt{m_e/m_i}xy\right]}{2 \sqrt{m_e/m_i} \gamma xy}
\end{eqnarray}
we find
\begin{eqnarray}
& &\frac{d T_i}{dt}=-\frac{8}{3}\frac{e^4 m_e^{3/2} e^{-\mu}}{\pi^{3/2} \hbar^3 \beta_e \sqrt{m_i}} \int_0^{\infty} \frac{dy}{y^2} \int_{-\infty}^{\infty} dx x \frac{e^{-x^2}}{|\epsilon(x,y)|^2}\cr
&\times& \sinh\left[2(\gamma-1)\sqrt{m_e/m_i}xy\right] \frac{e^{-(m_e/m_i+\gamma)y^2}}{1+e^{-\mu-\gamma y^2}} \ . \label{eq:integral_sinh}
\end{eqnarray}
An interesting thing to note here is that if we drop all quantum diffraction terms and set $\epsilon(x,y)=1$ in the previous integral, the result is exactly Brysk's correction to the Landau-Spitzer formula. The details of this are given in Appendix \ref{sec:Brysk}.

To facilitate our later treatment of the dielectric function, we make use of the formula
\begin{equation}
 \frac{1}{|\epsilon(x,y)|^2}=\frac{1}{2 i \mathrm{Im} \epsilon(x,y)} \left[\frac{1}{\epsilon^*(x,y)}-\frac{1}{\epsilon(x,y)} \right] \label{eq:epsdecomp}
\end{equation}
and note that $\epsilon^*(x,y)=\epsilon(-x,y)$. Similarly $\mathrm{Im}\epsilon(-x,y)=-\mathrm{Im}\epsilon(x,y)$. We can use these in the first term in (\ref{eq:epsdecomp}), and then change the integration variable $x \rightarrow -x$ to see that the replacement
\begin{equation}
 \frac{1}{|\epsilon(x,y)|^2} \rightarrow - \frac{1}{i} \frac{1}{\mathrm{Im} \epsilon(x,y)}\frac{1}{\epsilon(x,y)}
\end{equation}
does not change the integral. We also expand the $\sinh$ in the integrand
\begin{equation}
 \sinh\left[2(\gamma-1)\sqrt{m_e/m_i}xy\right]=\sum_{n=0}^{\infty}  \frac{\rho^{2n+1}}{(2n+1)!}y^{2n+1}x^{2n+1}
\end{equation}
where $\rho$ is defined in (\ref{eq:defs}). In general, only one or two $n$ need to be retained.

The integral is now
\begin{eqnarray}
& &\frac{d T_i}{dt}=\frac{8}{3i}\frac{e^4 m_e^{3/2} e^{-\mu}}{\pi^{3/2} \hbar^3 \beta_e \sqrt{m_i}} \sum_{n=0}^{\infty} \frac{\rho^{2n+1}}{(2n+1)!} \cr
&\times& \int_0^{\infty} dy \int_{-\infty}^{\infty} dx \frac{x^{2n+2} y^{2n-1}}{\mathrm{Im} \epsilon(x,y)} \frac{e^{-x^2}}{\epsilon(x,y)} \frac{e^{-(m_e/m_i+\gamma)y^2}}{1+e^{-\mu-\gamma y^2}} \ . \label{eq:integral}
\end{eqnarray}
It appears we have made this quantity complex, but the real part of the integral is zero by symmetry, leaving it purely imaginary to cancel the $i$ in the prefactor. For the coming work, this is the most useful form.

\section{Exact solution for $\alpha=0$} \label{sec:alphazero}
Given the smallness of $\alpha$, it is very tempting just to set $\alpha=0$ in the electron response function. We will not resist this temptation, at least for the moment. It may make one nervous, however, primarily because sum rules no longer produce the correct results. On the other hand, it seems that this is not really a problem outside of the coupled mode regime, as we will show in section \ref{sec:numerical}. It has been previously pointed out that this approximation allows an exact evaluation of integrals similar to (\ref{eq:dTdti}) in the context of the conductivity problem \cite{Oberman, Williams} although our method is new, as far as we know.

Setting $\alpha=0$ in the dielectric function leaves us with
\begin{eqnarray}
 \epsilon(x,y)=1+\frac{\eta^2}{y^2}\left[\frac{\gamma_{\mathrm{eff}}}{Z}+1-2xF(x)+i\sqrt{\pi}e^{-x^2}\right] \label{eq:epsalphazero}
\end{eqnarray}
and we have the simplification
\begin{equation}
\frac{x e^{-x^2}}{\mathrm{Im}\epsilon(x,y)}=\frac{y^2}{\eta^2}\frac{1}{\sqrt{\pi}}.
\end{equation}
The main observation that aids the calculation is that in the complex plane, for large complex $x \rightarrow z$, we have the Laurent expansion
\begin{equation}
 \frac{\eta^2 \gamma_{\mathrm{eff}}}{y^2 Z \epsilon(z,y)}=\sum_{n=0}^{\infty} \frac{a_{2n}(y)}{z^{2n}} , \label{eq:epsexp}
\end{equation}
where the coefficients $a_{2n}(y)$ are calculated in Appendix \ref{sec:Laurent}. The $x$-integral we are planning to solve is
\begin{equation}
 I_x^n=\int_{-\infty}^{\infty} \frac{\eta^2 \gamma_{\mathrm{eff}}}{y^2 Z \epsilon(x,y)} x^{2n+1} dx .
\end{equation}
In the complex plane, we integrate along the $x$-axis, where there are no singularities, and close the path by integrating along the arc $|z|=R$, taking $R \rightarrow \infty$. We denote this arc component of the integration as $I_R$. Because there are no singularities in the upper half plane for the dielectric function (\ref{eq:epsalphazero}), we have
\begin{equation}
 I_x^n=-I_R^n ,
\end{equation}
where
\begin{equation}
 I_R^n = i \sum_{m=0}^\infty a_{2m}(y) \int z^{2 (n-m+1) \theta}d z .
\end{equation}
The integration is over the arc $z=R e^{i \theta}$ in the upper half plane, i.e. $\theta \in [0,\pi]$. This is easily performed, the result is in fact zero unless $m=n+1$, and we have
\begin{equation}
 I_x^n=-i \pi a_{2n+2}(y)
\end{equation}
so that
\begin{eqnarray}
& &\frac{d T_i}{dt}=-\frac{8}{3}\frac{Z^3 e^4 m_e^{3/2} e^{-\mu}}{\pi \hbar^3 \beta_e \sqrt{m_i} \eta^4 \gamma_{\mathrm{eff}}} \sum_{n=0}^{\infty} \frac{\rho^{2n+1}}{(2n+1)!} J_n  \label{eq:integral2}
\end{eqnarray}
where
\begin{equation}
 J_n \equiv \int_0^{\infty} y^{2n+3} \frac{e^{-(m_e/m_i+\gamma)y^2}}{1+e^{-\mu-\gamma y^2}} a_{2n+2}(y) dy . \label{eq:Jn}
\end{equation}
It is sufficient for a wide range of conditions to calculate these for $n=0$ and $1$. Starting with $n=0$, we have, after a convenient change of variables
\begin{equation}
 J_0=\frac{\gamma_{\mathrm{eff}}\eta^4}{4 Z}\int_0^\infty \frac{te^{-t}}{(1+e^{-\mu-rt})(t+1/\Lambda)^2}dt \label{eq:J0}
\end{equation}
where
\begin{equation}
\frac{1}{\Lambda} \equiv \frac{\gamma_{\mathrm{eff}}\eta^2(m_e/m_i+\gamma)}{Z}
\end{equation}
which is generally a small quantity in weak coupling. The parameter $r \equiv \gamma/(m_e/m_i+\gamma)$ we refer to as the Brysk number, for the following reason. When $\gamma \gg m_e/m_i$, which is to say for essentially all conditions of interest, $r \approx 1$. However, in the opposite limit, when $\gamma \ll m_e/m_i$, $r \approx 0$ and we can take the first factor in the denominator of the integrand in (\ref{eq:J0}) outside of the integral, so quantum degeneracy just produces a Brysk multiplicative correction. Normally this correction is somewhat useful at weak degeneracy, but when $r \approx 0$, it is exactly the right thing to do. This is of course a rare situation, where we are unlikely to apply this formula anyway, so we will set $r=1$ from now on.

We now have
\begin{equation}
 J_0=\frac{\gamma_{\mathrm{eff}}\eta^4}{4 Z} \tilde{f}\left(\frac{1}{\Lambda}\right)
\end{equation}
where
\begin{equation}
 \tilde{f}(x) \equiv \int_0^\infty \frac{te^{-t}}{(1+e^{-\mu-t})(t+x)^2}dt \ . \label{eq:ftilde}
\end{equation}
This is a special function that does not appear to be expressible in terms of anything simple. Integrating it numerically would certainly not prove to be much of a challenge but we really only need to be able to evaluate it for small $x$. In Appendix \ref{sec:special_func}, we derive the needed expansion, which is a somewhat tricky procedure. To order $x^3 \ln x$ (we also have some pieces of higher-order terms),
\begin{eqnarray}
& &\tilde{f}(x)=e^x \left[U_1(\mueff)-\frac{e^{\mueff}}{B(\mueff)}-\frac{e^{\mueff}}{B(\mueff)}\ln x  \right. \cr
&-&\frac{e^{2 \mueff}}{[B(\mueff)]^2}x \ln x+\left(\frac{2e^{2\mueff}}{[B(\mueff)]^2}-U_2(\mueff) \right)x \cr
&+&\left. \frac{e^{2 \mueff}(-1+e^{\mueff})}{4 [B(\mueff)]^3}x^2-\frac{e^{2\mueff}(1-4e^{\mueff}+e^{2 \mueff})}{36 [B(\mueff)]^4}x^3 \right] \cr
& & \label{eq:fser}
\end{eqnarray}
where $B(\mu)$ is part of the Brysk degeneracy factor,
\begin{equation}
 B(\mu) \equiv 1+e^\mu \label{eq:B}
\end{equation}
and $\mueff$ is an effective chemical potential,
\begin{equation}
 \mueff=\mu-x .
\end{equation}
The motivation for the latter is given in Appendix \ref{sec:special_func}. The numbers $U_1(\mu)$ and $U_2(\mu)$ are defined by
\begin{eqnarray}
 U_1(\mu) &\equiv& \int_0^\infty \frac{\ln t e^{-t}}{(1+e^{-\mu-t})^2} dt \\ \label{eq:C}
 U_2(\mu) &\equiv& -\int_0^\infty \frac{\ln t (e^{-t}-e^{-2t-\mu})}{(1+e^{-\mu-t})^3} dt . \label{eq:C3}
\end{eqnarray}
In the classical limit, these are $U_1(\mu) \rightarrow \gamma_E$ and $U_2(\mu) \rightarrow - \gamma_E$, where $\gamma_E \approx 0.57722$ is the Euler constant. To evaluate these functions, we use the fits,
\begin{equation}
U_1(\mu)=
 \begin{cases}
 0.949714 e^{\mu} \ln |\mu| & \mu < -5 \\
 a_0+a_1 \mu+a_2 \mu^2+a_3 \mu^3 & \\
 \ \ \ \ \ \ \  +a_4 \mu^4  + a_5 \mu^5+a_6 \mu^6 &  -5 \le \mu \le 1 \\
  -\gamma_E \tanh[0.4753(\mu+0.04989)] & \mu > 1
 \end{cases}
\end{equation}
\begin{equation}
U_2(\mu)=
 \begin{cases}
 1.16511 e^{\mu}/\mu & \mu < -4 \\
 a_0+a_1 \mu+a_2 \mu^2+a_3 \mu^3 & \\
 \ \ \ \ \ \ \  +a_4 \mu^4  + a_5 \mu^5+a_6 \mu^6 &  -4 \le \mu \le 1 \\
  \gamma_E \tanh[0.4914(\mu-0.772571)] & \mu > 1
 \end{cases}
\end{equation}
Putting all these things together we have, for $n=0$,
\begin{equation}
 \frac{dT_i}{dt}^{(0)}=-\frac{4}{3}\frac{e^4 Z^2 m_e^{2} e^{-\mu}}{\pi \hbar^3 \beta_e m_i}(\gamma-1) \tilde{f}\left(\frac{1}{\Lambda}\right) . \label{eq:main}
\end{equation}
Although this is sufficient for many applications, we will also add the correction for $n=1$.

To do this, we need $J_1$ from (\ref{eq:Jn}). This leads to yet more special functions for which we again need the small $x$ expansions. The procedure is essentially no different from what we have already shown so we omit the derivations. The result is
\begin{eqnarray}
\frac{dT_i}{dt}^{(1)}&=&-\frac{4}{9}\frac{e^4 Z^2 m_e^{3} e^{-\mu}}{\pi \hbar^3 \beta_e m_i^2}(\gamma-1)^3 \cr
 &\times& \left[\eta^2 \tilde{f}_5 \left(\frac{1}{\Lambda}\right)+\frac{3}{\gamma+m_e/m_i}\tilde{f}_4 \left(\frac{1}{\Lambda}\right)\right] \label{eq:main1}
\end{eqnarray}
where
\begin{eqnarray}
& &\tilde{f}_4(x)=e^x \bigg[ e^{\mueff} \ln (1+e^{-\mueff})-2 U_1(\mueff) x \cr
&+&  \left(U_2(\mueff)-\frac{5 e^{2 \mueff}}{2 [B(\mueff)]^2} \right) x^2 + \frac{e^{2 \mueff}(1-e^{\mueff})}{6 [B(\mueff)]^3}x^3 \cr
&+& \frac{2 e^{\mueff}}{B(\mueff)}x \ln x +\frac{e^{2 \mueff}}{[B(\mueff)]^2} x^2 \ln x \bigg],
\end{eqnarray}
\begin{eqnarray}
& &\tilde{f}_5(x)= e^x \bigg[-\frac{3 e^{\mueff}}{2 B(\mueff)} + U_1(\mueff) \cr
&+& \left(\frac{2 e^{2 \mueff}}{[B(\mueff)]^2}-2 U_2(\mueff) \right) x  \cr 
&+& \left(\frac{3 e^{2 \mueff}(e^{\mueff}-1)}{2 [B(\mueff)]^3} + U_3 (\mueff) \right) x^2 \cr
&+& \frac{e^{2 \mueff}(1-4 e^{\mueff}+e^{2 \mueff})}{18 [B(\mueff)]^4}x^3 - \frac{e^{\mueff}}{B(\mueff)} \ln x \cr
&+& \frac{e^{2 \mueff}(1-e^{2 \mueff})}{2 [B(\mueff)]^3}x^2 \ln x \bigg]
\end{eqnarray}
and
\begin{equation}
 U_3(\mu) \equiv \frac{e^{-2\mu}}{2} \int_0^{\infty} \frac{e^{-3t}(1-4 e^{\mu+t}+e^{2 \mu+2t}) \ln t}{(1+e^{-\mu-t})^4}dt .
\end{equation}
Once again, we use a fit for $U_3(\mu)$,
\begin{equation}
U_3(\mu)=
 \begin{cases}
 -1.01714 e^{\mu}/(2 \mu^2) & \mu < -3.75 \\
 a_0+a_1 \mu+a_2 \mu^2+a_3 \mu^3 & \\
 \ \ \ \ \ \ \  +a_4 \mu^4  + a_5 \mu^5+a_6 \mu^6 &  -3.75 \le \mu \le 1 \\
  -0.5 \gamma_E \tanh[0.5241(\mu-1.6374)] & \mu > 1
 \end{cases}
\end{equation}
with the coefficients given in Table \ref{tab:fit}. Equation (\ref{eq:main1}) is meant to be added to (\ref{eq:main}) if the temperature difference is large enough to require the next power in $\gamma-1$. This procedure can be carried on to arbitrary $n$, with the results becoming increasingly complicated, but the reader is left on his or her own for that; it is unclear that even (\ref{eq:main1}) is actually necessary for applications of current interest. We give some numerical examples in section \ref{sec:numerical}.

\begin{table}
\begin{center}
 \begin{tabular}{c|c c c}
  & $U_1$ & $U_2$ & $U_3$   \\
  \hline
$a_0$ & -0.0617725  & -0.118312    &  0.104306    \\
$a_1$ & -0.183813   &  0.0823933   &  0.0638929   \\
$a_2$ & -0.052559   &  0.0971156   & -0.0357814   \\
$a_3$ &  0.0183355  &  0.013315    & -0.033785    \\
$a_4$ &  0.0113972  & -0.00760402  & -0.00625476  \\
$a_5$ &  0.00199856 & -0.00246233  &  0.000446359 \\
$a_6$ &  0.00012039 & -0.000203598 &  0.000152431 \\
 \end{tabular}
 \caption{Fitting coefficients for the special functions $U_1(\mu)$, $U_2(\mu)$ and $U_3(\mu)$.}  \label{tab:fit}
\end{center}
\end{table}

\section{Coupled modes} \label{sec:coupledmode}
\begin{figure}
  \includegraphics[width=3in]{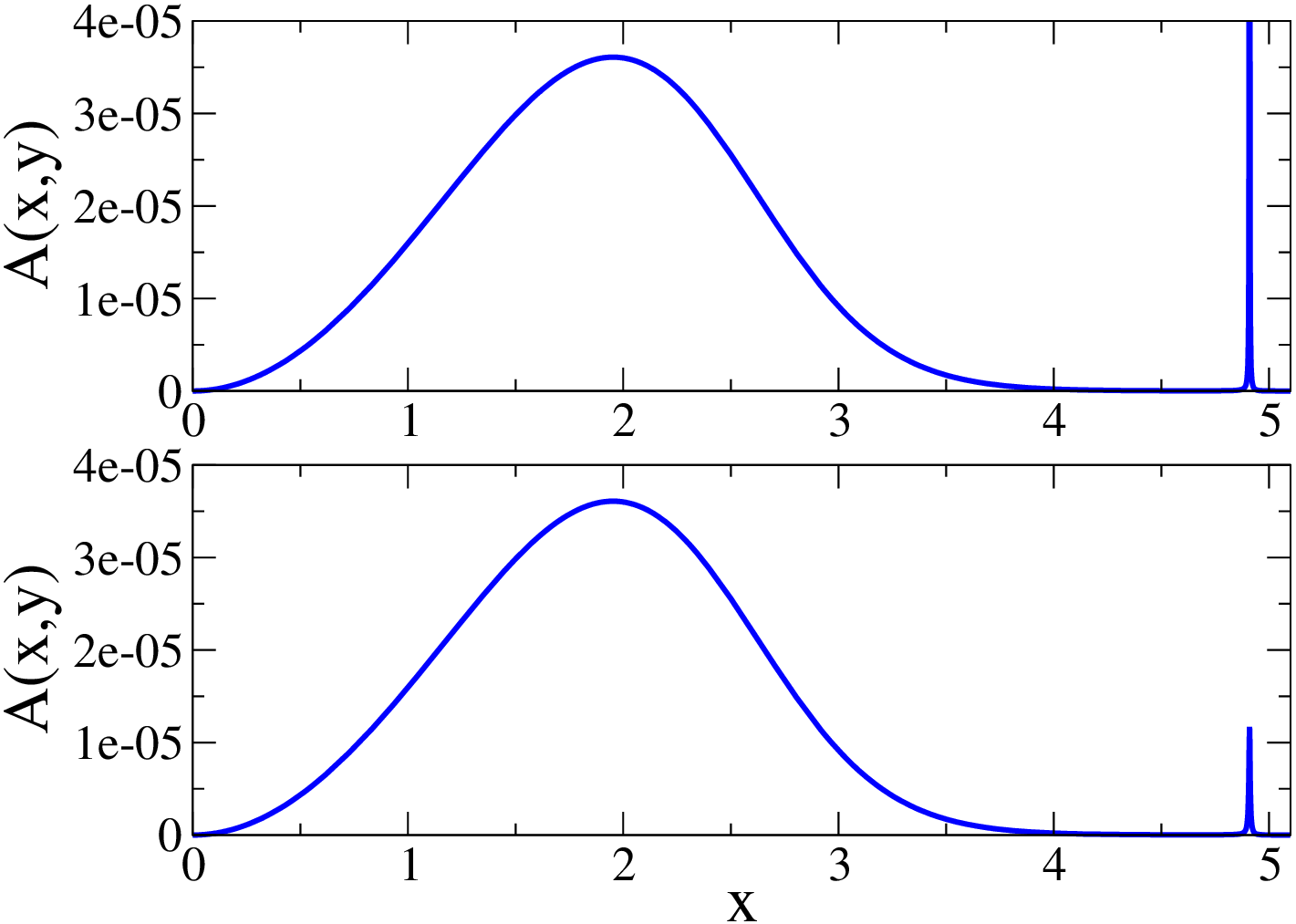}
 \caption{Top panel: plot of the piece (\ref{eq:Axy}) of the integrand (\ref{eq:integral_sinh}) for $y=2.1$ and $\alpha=0$; bottom panel: same as left but with $\alpha=1.8 \times 10^{-6}$. Retaining even a small $\alpha$ is crucial in getting the sharp ion-acoustic peak height correct.}
 \label{fig:alphacomp}
\end{figure}

Dropping $\alpha$ completely from the dielectric function, although leading to an accurate approximation for a wide range of conditions, does not allow us to capture the coupled mode effect completely. The crux of the problem is illustrated in Figure \ref{fig:alphacomp}, where we have plotted the piece of the integrand in equation (\ref{eq:integral_sinh}),
\begin{equation}
A(x,y) \equiv \frac{x e^{-x^2}}{|\epsilon(x,y)|^2}, \label{eq:Axy}
\end{equation}
for hydrogen at $T_i=1.0 \times 10^5 K$, $T_e = 3.0 \times 10^7 K$ and $n_e=n_i=10^{26} \mathrm{cm}^{-3}$ at $y=2.1$. At these conditions, $\gammaeff=0.0032$ and $\alpha=1.8 \times 10^{-6}$ . In the top panel in Figure \ref{fig:alphacomp}, we plot (\ref{eq:Axy}) for $y=2.1$ and $\alpha=0$ in the dielectric function and in the bottom panel is $A(x,2.1)$ with $\alpha$ retaining its physical value. As we can see, the two plots are qualitatively similar, each with a sharp ion acoustic peak around $x \approx 4.9$, but the height of the peak is far greater when $\alpha=0$. This is what causes the overestimation of equilibration rate in the coupled mode regime if we drop $\alpha$. Note, however, that we can always neglect $\alpha$ in the real part of the dielectric function because this piece primarily fixes the location of the peak; a small $\alpha$ will move it hardly at all. It is the imaginary part that determines the height, and here is where we need to be careful about dropping $\alpha$ in the coupled mode regime.

\begin{figure}
  \includegraphics[width=3in]{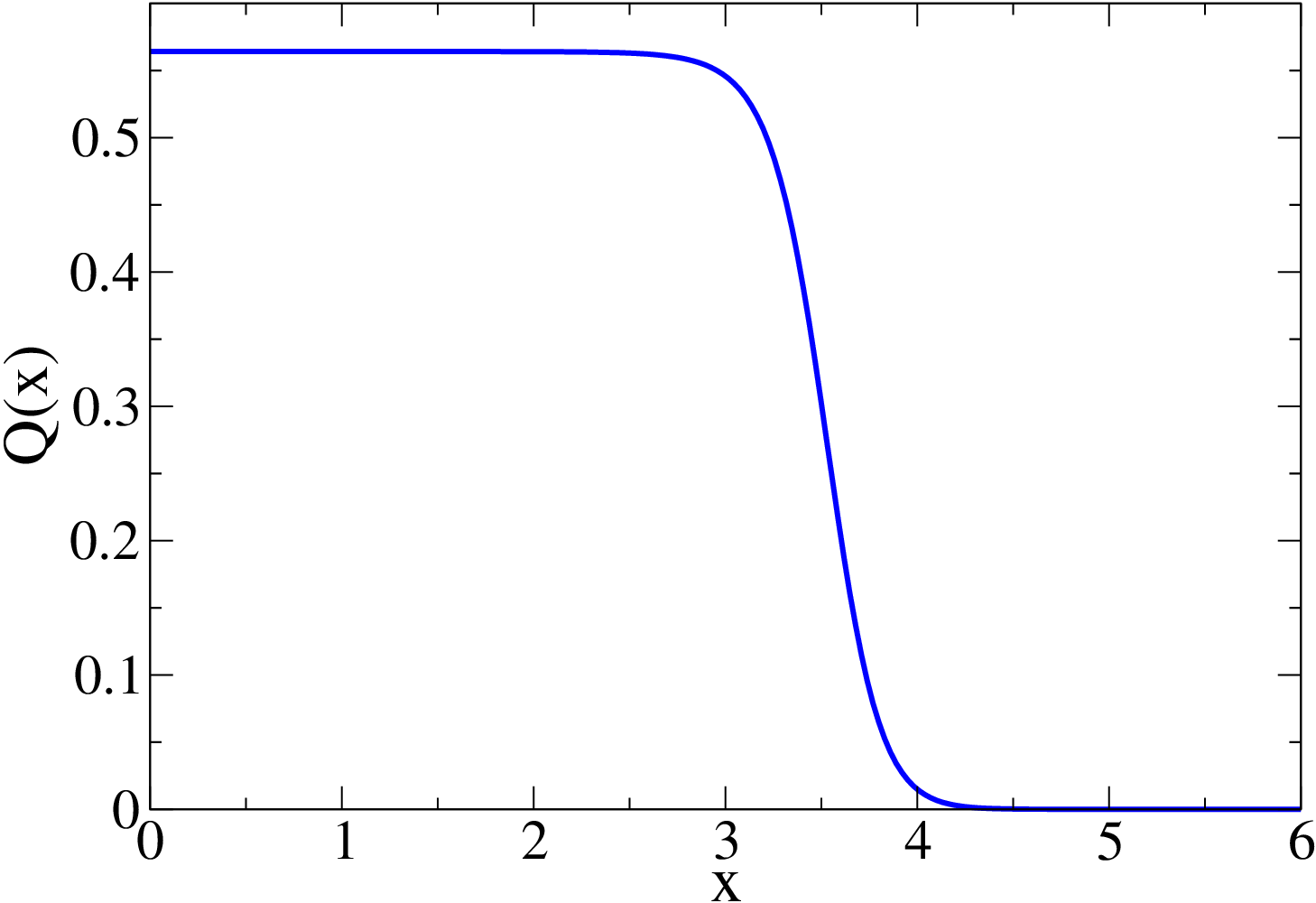}
 \caption{The function $Q(x)$ defined in (\ref{eq:Q}). This is constant if $\alpha=0$, but here $\alpha=1.8 \times 10^{-6}$ and $Q(x)$ is constant for $x \lesssim 3$ but then drops to zero. This function is a part of the integrand in (\ref{eq:Gamma}) and provides convergence in the coupled mode regime.}
 \label{fig:Q}
\end{figure}

Retaining $\alpha$ in the calculation of the previous section leads to complications that render the method impractical. Instead, we take the alternative approach of first integrating (\ref{eq:integral}) over the dimensionless wave number $y$. For this, we define the double integrals $\overline{J}_n$ by
\begin{eqnarray}
 & &\frac{d T_i}{dt}=\frac{8}{3i}\frac{e^4 m_e^{3/2}e^{-\mu}}{\pi^{3/2} \hbar^3 \beta_e \sqrt{m_i}} \sum_{n=0}^\infty \frac{\rho^{2n+1}}{(2n+1)!} \overline{J}_n \label{eq:dTdtCM} \\ 
\overline{J}_n &\equiv& \int_0^{\infty} \int_{-\infty}^{\infty} \frac{x^{2n+2} e^{-(1+\alpha)x^2}}{\eta^2 w_i(x)} \cr 
&\times& \frac{e^{-(m_e/m_i+\gamma)y^2}}{y^2+\eta^2 w(x)} \frac{y^{2n+3}}{1+e^{-(\alpha x^2+\gamma y^2 + \mu)}} dx dy\ . \label{eq:Jbar}
\end{eqnarray}
Now we expand the special function defined by the $y$-integral to a few orders in its argument, which in this case is a complex function. Isolating this $y$-integral, we define
\begin{equation}
 I_n[\eta^2 w(x)] \equiv \int_0^{\infty} \frac{y^{2n+3}}{1+e^{-(\gamma y^2 + \mu)}} \frac{e^{-(m_e/m_i+\gamma)y^2}}{y^2+\eta^2 w(x)}dy \ , \label{eq:In}
\end{equation}
where $w(x)=w_r(x)+i w_i(x)$ is given by (\ref{eq:wr}) and (\ref{eq:wi}). Clearly, the special function we need to study is
\begin{equation}
 f_n(z;\mu)\equiv \int_0^\infty \frac{e^{-t}t^{n+1}}{(t+z)(1+e^{-\mu-t})}dt
\end{equation}
Defining the variable $u$ to make the substitution
\begin{equation}
 u \equiv t+z
\end{equation}
to write
\begin{equation}
 f_n(z;\mu)=e^z \int_z^\infty \frac{e^{-u}(u-z)^{n+1}}{u(1+e^{-\mu+z-u})} du.
\end{equation}
Finding the expansion of this function in $z$ is tedious but straightforward. It follows a procedure similar to that outlined in Appendix \ref{sec:special_func} except that it is not possible to combine the $z$ inside the integrand into an effective chemical potential because $z$ is complex. For this, we must expand the integrand in powers of $z$ and then expand each of the resulting terms as is done in Appendix \ref{sec:special_func}. We omit these details, but the expansion is of the form
\begin{eqnarray}
 f_n(z;\mu)&\approx&f_{n}^{(0)}(\mu)+f_{n}^{(L)}(\mu)\log z+f_{n}^{(1L)}(\mu)z\log z \cr
 &+&f_{n}^{(2L)}(\mu)z^2 \log z + f_{n}^{(3L)}(\mu)z^3 \log z \cr
 &+&f_{n}^{(4L)}(\mu)z^4 \log z+f_{n}^{(1)}(\mu)z+f_{n}^{(2)}(\mu)z^2 \cr
 &+&f_{n}^{(3)}(\mu)z^3+f_{n}^{(4)}(\mu)z^4 \ . \label{eq:fnser}
\end{eqnarray}
It will turn out that we do not need the explicit forms of all of these coefficients. The only ones we do need are
\begin{eqnarray}
 f_{0}^{(1L)}(\mu)&=& f_{0}^{(2L)}(\mu)=\frac{e^{\mu}}{1+e^{\mu}} \\
 f_{0}^{(3L)}(\mu)&=& \frac{1}{2}f_{0}^{(1L)}(\mu) \\
 f_{1}^{(1L)}(\mu)&=& 0 \\
 f_{1}^{(2L)}(\mu)&=&f_{1}^{(3L)}(\mu)=-\frac{e^{\mu}}{1+e^{\mu}} .
\end{eqnarray}
Inserting our expansion of the $y$-integrand (\ref{eq:In}) into the integral (\ref{eq:integral}) leaves us with a one-dimensional integral over $x$ containing a complicated mixture of $w_r(x)$ and $w_i(x)$ resulting from inserting $z=\eta^2 w(x)$ into (\ref{eq:fnser}) and taking the imaginary part. The result is that we can expand the integrals $\overline{J}_n$ of equation (\ref{eq:Jbar}) as
\begin{eqnarray}
 \overline{J}_n &\approx& \frac{1}{\left(m_e/m_i+\gamma \right)^n} \left( \overline{J}_n^{(0)} + \overline{J}_n^{(L)}\log \phi+\overline{J}_n^{(1L)} \phi \ln \phi \right. \cr 
 & & \ \ \ \ \ \ + \overline{J}_n^{(2L)} \phi^2 \ln \phi + \overline{J}_n^{(3L)} \phi^3 \ln \phi + \overline{J}_n^{(1)} \phi \cr
 & & \ \ \ \ \ \ +\left. \overline{J}_n^{(2)} \phi^2 + \overline{J}_n^{(3)} \phi^3 \right) , \label{eq:Jnbarexp}
\end{eqnarray}
where each $\overline{J}$ is an integral over $x$ and
\begin{equation}
 \phi \equiv (\gamma+m_e/m_i) \eta^2
\end{equation}
is an expansion parameter that serves the same purpose as $1/\Lambda$ in section \ref{sec:alphazero}. This definition is more convenient for the coupled mode calculations. Each integral in (\ref{eq:Jnbarexp}) is a function of $\alpha$ and $\gammaeff/Z$ and if we set $\alpha=0$ in all of these, the result should be identical with (\ref{eq:main}). There are in fact not many terms in (\ref{eq:Jnbarexp}) that are sensitive to setting $\alpha=0$; the only ones that matter are of the form $\overline{J}_n^{(j)}$, i.e., the terms that do not involve $\log \phi$. Within these terms, we have the integrals
\begin{eqnarray}
& &\Gamma_n^{(j)}\left(\frac{\gammaeff}{Z},\alpha \right) \cr
& &\equiv \int_{-\infty}^\infty e^{-x^2}x^{2n+2} \frac{w_r^{j+1}(x)}{w_i(x)} \arctan [w_r(x),w_i(x)] dx \label{eq:Gamma}
\end{eqnarray}
where $\arctan(x,y)$ is the four-quadrant version of $\tan^{-1} y/x$. The arctangent arises from the logarithmic terms in the series (\ref{eq:fnser}) because for complex $w(x)$,
\begin{equation}
 \ln w(x)=\ln |w(x)|+i \theta
\end{equation}
where the angle $\theta$ is given by the arctangent. It is interesting to consider how exactly the integral (\ref{eq:Gamma}) converges. First, if we have $\alpha=0$, the $w_i(x)$ in the denominator cancels $e^{-x^2}$ and convergence is left up to the arctan. Because $\tan^{-1} z \approx z$ for small $z$, one might think that the integrand goes to zero in the same manner as $w_i(x)$. This is essentially correct, but if $w_r(x)$ is negative, then $\arctan [w_r(x),w_i(x)]$ goes to $\pi$, no matter how small $w_i(x)$ becomes, and the integrand cannot be zero until $w_r(x)$ becomes positive again. When do we have to worry about $w_r(x)$ being negative? This happens when $\gammaeff/Z$ is sufficiently small, and as we can see from equation (\ref{eq:wr}), $w_r(x)$ is always positive provided
\begin{equation}
 \frac{\gamma_{\mathrm{eff}}}{Z}> | \min (1-2x F(x)) | \approx 0.28,
\end{equation}
hence the condition (\ref{eq:CMcond}). The smaller $\gammaeff/Z$, the larger the $x$ at which $w_r(x)$ becomes positive again, and thus the larger the integral. If, however, we have a non-zero $\alpha$ then the factor in the integrand,
\begin{equation}
 Q(x) \equiv \frac{x e^{-x^2}}{w_i(x)} , \label{eq:Q}
\end{equation}
which is constant if $\alpha=0$, provides its own mode of convergence if $\gammaeff/Z$ is very small. In Figure \ref{fig:Q}, we plot $Q(x)$ for $\alpha=1.8 \times 10^{-6}$. It is constant for $x \lesssim 3$ but then falls to zero, providing an earlier cutoff than the arctangent if $\gammaeff/Z$ is sufficiently small.

The point of this discussion is that when coupled modes are important we should correct equation (\ref{eq:main}) by subtracting the piece containing the integral
\begin{equation}
\overline{\Gamma}(\gammaeff/Z)\equiv \Gamma(\gammaeff/Z,0)
\end{equation}
and adding $\Gamma(\gammaeff/Z,\alpha)$. This correction then looks like
\begin{eqnarray}
& &\Delta R \equiv \frac{4}{3} \frac{e^4 m_e^{3/2} e^{-\mu}}{\pi^{3/2} \hbar^3 \beta_e \sqrt{m_i}} \sum_{n=0}^{\infty} \frac{\rho^{2n+1}}{(2n+1)!} \cr
&\times& \bigg[ \left( \Gamma_n^{(0)}(\gammaeff/Z,\alpha )-\overline{\Gamma}_n^{(0)}(\gammaeff/Z) \right) f_n^{(1L)}(\mu)\cr
&+&\left( \Gamma_n^{(1)}(\gammaeff/Z,\alpha )-\overline{\Gamma}_n^{(1)}(\gammaeff/Z) \right) f_n^{(2L)}(\mu) \phi \cr
&+&\left( \Gamma_n^{(2)}(\gammaeff/Z,\alpha )-\overline{\Gamma}_n^{(2)}(\gammaeff/Z) \right) f_n^{(3L)}(\mu) \phi^2 \bigg]. \label{eq:DeltaR}
\end{eqnarray}
The remaining question is how to evaluate the $\Gamma$ integrals. Starting with $\overline{\Gamma}_n^{(j)}(\gammaeff/Z)$, these are functions of only a single variable and power series can be derived for them; they are given in Appendix \ref{sec:Gammaseries}. As for $\Gamma_n^{(j)}(\gammaeff/Z,\alpha)$, we could also try a series or a fit, but instead we will just use a simple 10-point Gaussian quadrature. Because of the weight $e^{-x^2}$ in (\ref{eq:Gamma}) and the fact that the integrand is even, we make the substitution $u=x^2$ and use a Gauss-Laguerre scheme. The integral is then approximated by
\begin{equation}
 \Gamma_n^{(j)}\left(\frac{\gammaeff}{Z},\alpha \right) \approx \sum_{i=1}^{N} W_i g^{(j)}_n(u_i)
\end{equation}
where $u_i$ are the zeros of the the $N^{\mathrm{th}}$ associated Laguerre polynomial of order $1/2$, $L_{N}^{(1/2)}(u)$, $W_i$ are the weights, given by
\begin{equation}
 W_j=\frac{u_j \Gamma(N+1/2)}{N! (N+1/2)\left[L_{N-1}^{(1/2)}(u_j)\right]^2} \ , \label{eq:quad}
\end{equation}
and $g^{(j)}_n(u)$ is the part of integrand of (\ref{eq:Gamma}) not including the factor $u^{1/2}e^{-u}$,
\begin{equation}
 g^{(j)}_n(u) \equiv u^n \frac{w_r^{j+1}(\sqrt{u})}{w_i(\sqrt{u})} \arctan [w_r(\sqrt{u}),w_i(\sqrt{u})]
\end{equation}
The weights, $W_i$, and the abscissa points, $u_i$, are given in Table \ref{table:constants} for $N=10$. The function $g(u)$ must be computed at the points $u_i$, but this is readily accomplished since no special functions need to be evaluated; the Dawson function in (\ref{eq:wr}) can be precalculated at the points $u_i$, and these are given in Table \ref{table:constants} as $D_i = 2 \sqrt{u_i} F(\sqrt{u_i})$. 

This is all one needs to compute the correction to (\ref{eq:main}) given by (\ref{eq:DeltaR}) in the coupled mode regime, if necessary. As before, we include both the $n=0$ and $n=1$ terms but $n=0$ should be sufficient for most applications.
\begin{center}
\begin{table}
 \begin{tabular}{|c|c|c|c|c|}
 \hline 
 j & $W_j$ & $u_j$ & $D_j$ & $\sqrt{u_j}$ \\
 \hline
 1 & 0.17547082 & 0.22987298 & 0.39536421 & 0.47945071 \\
 2 & 0.35522339 & 0.92448155 & 1.03900923 & 0.96149963 \\
 3 & 0.25268356 & 2.09941046 & 1.28306187 & 1.44893425 \\
 4 & 0.08635610 & 3.78288087 & 1.21745057 & 1.94496295 \\
 5 & 0.01510978 & 6.01991803 & 1.12171083 & 2.45355212 \\
 6 & 0.00132822 & 8.88034760 & 1.07086902 & 2.97999121 \\
 7 & 0.00005419 & 12.4748324 & 1.04633928 & 3.53197288 \\
 8 & $8.73747587 \times 10^{-7}$ & 16.9908473 & 1.03251942 & 4.12199555 \\
 9 & $4.01969989 \times 10^{-9}$ & 22.7910029 & 1.02357156 & 4.77399234 \\
 10 & $2.29222153 \times 10^{-12}$ & 30.8064059 & 1.01709339 & 5.55035187 \\
 \hline
 \end{tabular}
\caption{Constants used in the quadrature scheme; the weights, $W_i$; the abscissa points $u_i$; the Dawson function evaluated at the abscissa points $D_i=2 \sqrt{u_i} F(\sqrt{u_i})$; the square roots of $u_i$.}
 \label{table:constants}
\end{table}
\end{center}

\section{Numerical examples} \label{sec:numerical}
\begin{figure}
  \includegraphics[width=3.25in]{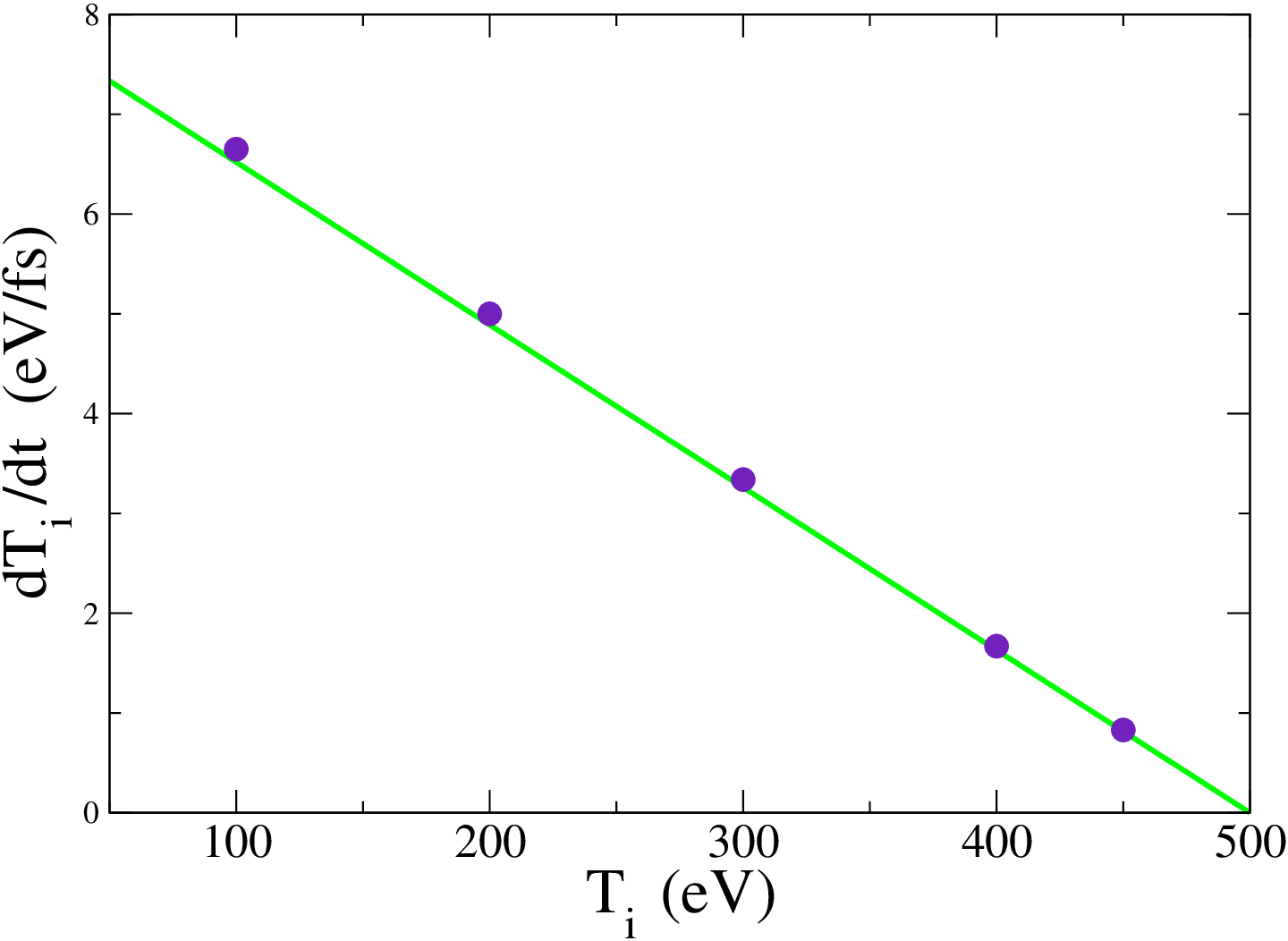}
 \caption{Comparison of (\ref{eq:main}) (line) with a numerical evaluation of the double integral (\ref{eq:dTdti}) (circles) for hydrogen ($Z$=1) at $n_i=n_e=10^{26}$ and $T_e=500 \mathrm{eV}$. Fermi energy: $E_F=786.60$ eV.}
 \label{fig:comp1}
\end{figure}

\begin{figure}
  \includegraphics[width=3.25in]{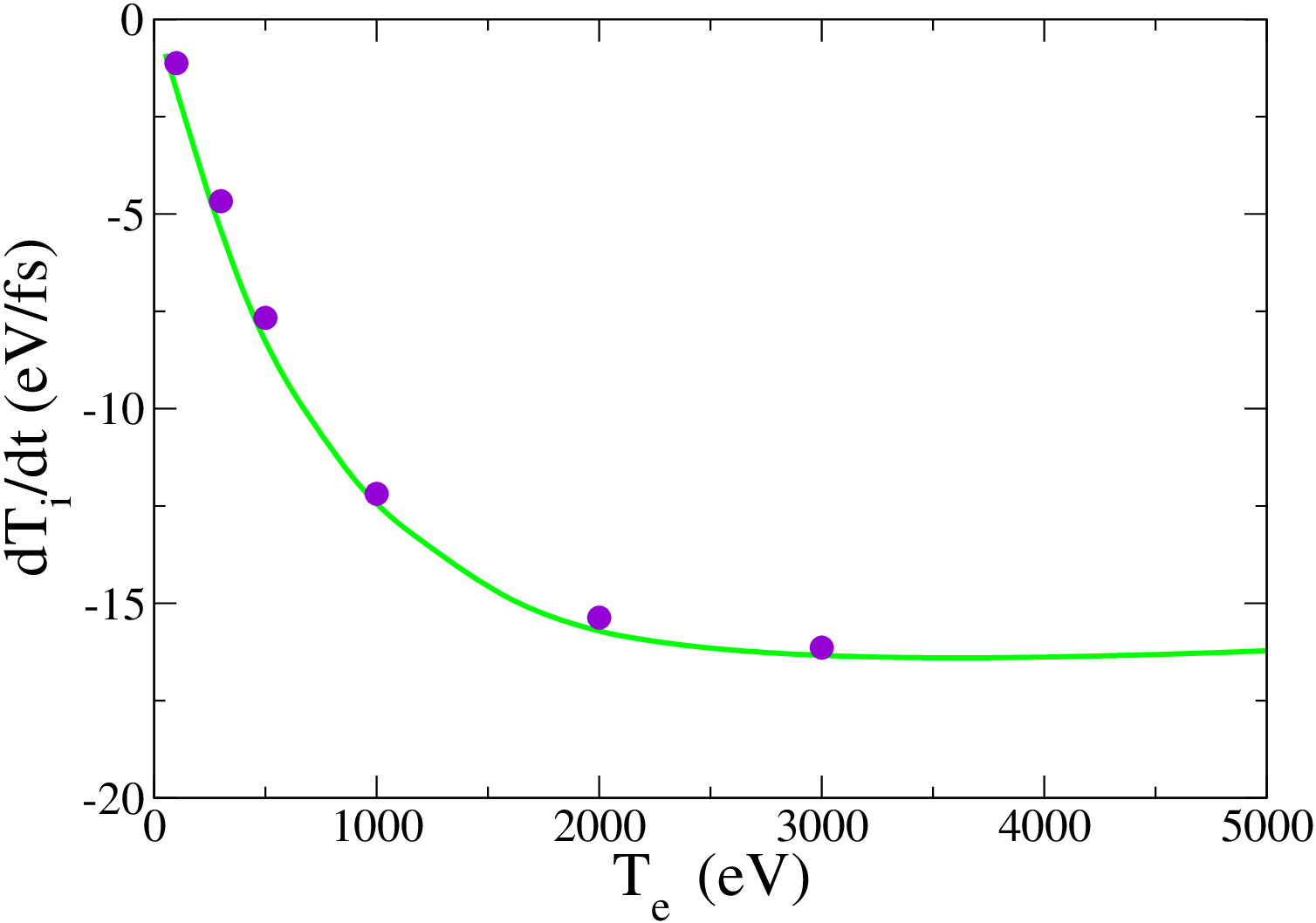}
 \caption{Comparison of (\ref{eq:main}) (line) with a numerical evaluation of the double integral (\ref{eq:dTdti}) (circles) for argon ($Z$=18) at $n_i=n_e/Z=10^{25}$ and $T_i=1.1 T_e \mathrm{eV}$ for a range of $T_e$. Fermi energy: $E_F=1162.48$ eV.}
 \label{fig:comp2}
\end{figure}

Here, we compare our formula (\ref{eq:main}) to direct numerical integrations of (\ref{eq:dTdti}) in which we neglect neither $\alpha$ nor quantum diffraction in the dielectric function. In Figure \ref{fig:comp1}, we give some example calculations for hydrogen at $n=10^{26} \mathrm{cm}^{-3}$, $T_e=500 \mathrm{eV}$ and various ion temperatures. The agreement between our formula and the numerical integration is nearly perfect. In Figure \ref{fig:comp2}, we show the case of argon ($Z=18$) at $n_i=10^{25} \mathrm{cm}^{-3}$ over a range of $T_e$ with $T_i=1.1 T_e$. This plot covers a wide range of electron degeneracy and once again the agreement with the full integration is very good. These results are typical of the performance of (\ref{eq:main}) over a wide range of conditions of practical interest.

Next, we examine the coupled mode correction, equation (\ref{eq:DeltaR}). Table \ref{table:CM} gives various numerical examples, using only $n=0$ of equation (\ref{eq:DeltaR}), at conditions where the coupled mode effect is expected to be important. We include here calculations done with the Fermi golden rule (FGR) approximation, equation (\ref{eq:FGR}), which can be used at non-degenerate conditions. We find that (\ref{eq:DeltaR}) does a good job of correcting (\ref{eq:main}) to capture the coupled mode effect. One interesting thing to note here is that the FGR results are numerically very close to (\ref{eq:main}). It is not completely obvious that this should be the case, as we have nowhere assumed the factorization (\ref{eq:FGR}). As illustrated clearly in Ref.  \cite{Vorberger}, the FGR approximation both moves the position of the ion-acoustic pole and alters its height. In contrast, as shown in Figure \ref{fig:alphacomp}, in the coupled mode regime (\ref{eq:main}) does not correctly capture the height of the peak but at least locates it accurately. Apparently, this distinction is not important for the numerics at these conditions.

The small errors in Table \ref{table:CM} can be corrected by adding the next order term in $T_i-T_e$, equation (\ref{eq:main1}). In Table \ref{table:CM1} we show the result of adding (\ref{eq:main1}) to the coupled mode calculations. Obviously, this correction mostly accounts for the errors. However, they are quite small and correcting them is probably not important for practical applications, making (\ref{eq:main1}) of primarily academic interest. Adding the $n=1$ term from (\ref{eq:DeltaR}) changes the answer hardly at all for these conditions.

\begin{figure}
  \includegraphics[width=3.25in]{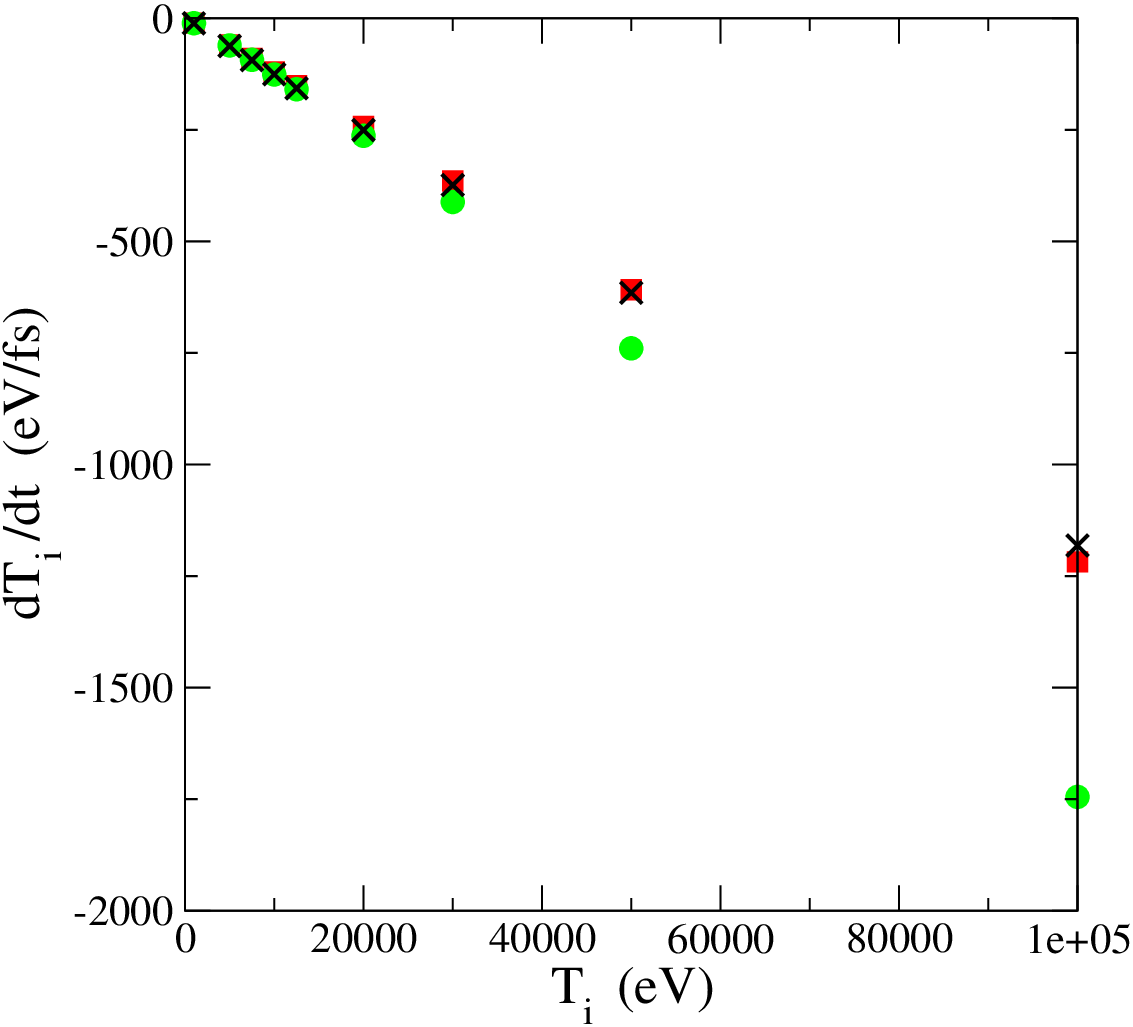}
 \caption{Equilibration rates calculated for hydrogen at $n_i=n_e=10^{25} \mathrm{cm}^{-3}$ and $T_e=100 \mathrm{eV}$ over a range of ion temperatures. Black x: direct integration of (\ref{eq:dTdti}); red squares: equation (\ref{eq:main}); green circles: equation (\ref{eq:main})+equation (\ref{eq:main1}). Fermi energy: $E_F=169.25$ eV.}
 \label{fig:hot_ions}
\end{figure}

So far, we have looked at cases for which the electron temperature is higher than the ion temperature. Equation (\ref{eq:main}) is also valid when the ions are hotter. In Figure \ref{fig:hot_ions} we show the rate computed for hydrogen at $n_i=n_e=10^{25} \mathrm{cm}^{-3}$, $T_e=100$ eV and a spread of ion temperatures. Over a wide range of temperature differences, equation (\ref{eq:main}) provides an excellent approximation. We also plot the correction (\ref{eq:main1}) and we can see that it does provide the required, but miniscule, correction at lower ion temperatures but at very large ion temperatures, where (\ref{eq:main}) begins breaking down, (\ref{eq:main1}) does not make things any more accurate. The reason for this is that we have discarded $\alpha$ in several places, both inside and outside of the dielectric function, and when $\alpha$ starts to become large, as it will when $T_i \gg T_e$, there is no reason to believe that either (\ref{eq:main}) or its correction via (\ref{eq:main1}) will provide an accurate estimate of the integral. Evidently, according to Figure \ref{fig:hot_ions}, when this occurs one is better off simply using (\ref{eq:main}) on its own. Once again, this is probably not of much practical concern.

\begin{center}
\begin{table*}
 \begin{tabular}{|c|c|c|c|c|c|c|c|}
 \hline 
 $n_i$ & $T_i$ (K) & $T_e$ (K) & eq.(\ref{eq:main}) (eV/fs) & eq.(\ref{eq:main})+eq.(\ref{eq:DeltaR}) & FGR (eV/fs) & eq.(\ref{eq:dTdti})  \\
 \hline
 $1 \times 10^{23}$ & $1 \times 10^5$ & $3 \times 10^7$ & 0.0346 & 0.0292 & 0.0350 & 0.0297 \\
 $1 \times 10^{24}$ & $1 \times 10^5$ & $3 \times 10^7$ & 0.274  & 0.220 & 0.279  & 0.225  \\
 $1 \times 10^{25}$ & $1 \times 10^5$ & $3 \times 10^7$ & 2.017  & 1.49   & 2.06   & 1.54   \\
 \hline
 \end{tabular}
\caption{Comparison of formulas with numerical evaluations of (\ref{eq:dTdti}) and the Fermi golden rule (FGR) in the coupled mode regime. Equation (\ref{eq:main}) on its own does not capture the coupled mode effect but rather closely matches FGR. Adding the correction (\ref{eq:DeltaR}) brings the results much closer to their coupled mode values.}
 \label{table:CM}
\end{table*}
\end{center}
\begin{center}
\begin{table*}
 \begin{tabular}{|c|c|c|c|c|c|c|c|}
 \hline 
 $n_i$ & $T_i$ (K) & $T_e$ (K) & eq.(\ref{eq:main})+eq.(\ref{eq:main1}) (eV/fs) & eq.(\ref{eq:main})+eq.(\ref{eq:main1})+eq.(\ref{eq:DeltaR}) & FGR (eV/fs) & eq.(\ref{eq:dTdti})  \\
 \hline
 $1 \times 10^{23}$ & $1 \times 10^5$ & $3 \times 10^7$ & 0.0350 & 0.0296 & 0.0350 & 0.0297 \\
 $1 \times 10^{24}$ & $1 \times 10^5$ & $3 \times 10^7$ & 0.278  & 0.225 & 0.279  & 0.225  \\
 $1 \times 10^{25}$ & $1 \times 10^5$ & $3 \times 10^7$ & 2.07  & 1.54   & 2.06   & 1.54   \\
 \hline
 \end{tabular}
\caption{The effect of adding the large-temperature correction, equation (\ref{eq:main1}). Adding it only to (\ref{eq:main}) brings the result into essentially exact agreement with the FGR, whereas combining it with the correction (\ref{eq:DeltaR}) gives an answer nearly indistinguishable from the numerical coupled mode result.}
 \label{table:CM1}
\end{table*}
\end{center}

\section{Conclusion}
We have derived an analytic expression for the electron-ion temperature equilibration rate predicted by the Lenard-Balescu integral, (\ref{eq:dTdti}). The main result, equation (\ref{eq:main}), closely matches numerical integrations of (\ref{eq:dTdti}) over most conditions of practical interest, is valid for arbitary electron degeneracy, and is suitable for fast computations within a larger simulation. We also include corrections for the coupled mode effect and for large temperature differences. However, it is likely that for most practical applications equation (\ref{eq:main}) is perfectly sufficient without these corrections. Our method for exactly solving dielectric function integrals, namely the Laurent expansion for large complex frequency, can probably be applied to computing other properties for which Lenard-Balescu integrals appear, such as thermal and electrical conductivities \cite{Williams,Whitley}.


\section*{Acknowledgements}
Susana Serna was supported by Spanish MINECO grant MTM2014-56218-C2-2-P. This work was performed under the auspices of the U.S. Department of Energy at the Lawrence Livermore National Laboratory under Contract No. DE-AC52-07NA27344.

\bibliography{T_equilibration.bib}

\begin{thebibliography}{22}%
\makeatletter
\providecommand \@ifxundefined [1]{%
 \@ifx{#1\undefined}
}%
\providecommand \@ifnum [1]{%
 \ifnum #1\expandafter \@firstoftwo
 \else \expandafter \@secondoftwo
 \fi
}%
\providecommand \@ifx [1]{%
 \ifx #1\expandafter \@firstoftwo
 \else \expandafter \@secondoftwo
 \fi
}%
\providecommand \natexlab [1]{#1}%
\providecommand \enquote  [1]{``#1''}%
\providecommand \bibnamefont  [1]{#1}%
\providecommand \bibfnamefont [1]{#1}%
\providecommand \citenamefont [1]{#1}%
\providecommand \href@noop [0]{\@secondoftwo}%
\providecommand \href [0]{\begingroup \@sanitize@url \@href}%
\providecommand \@href[1]{\@@startlink{#1}\@@href}%
\providecommand \@@href[1]{\endgroup#1\@@endlink}%
\providecommand \@sanitize@url [0]{\catcode `\\12\catcode `\$12\catcode
  `\&12\catcode `\#12\catcode `\^12\catcode `\_12\catcode `\%12\relax}%
\providecommand \@@startlink[1]{}%
\providecommand \@@endlink[0]{}%
\providecommand \url  [0]{\begingroup\@sanitize@url \@url }%
\providecommand \@url [1]{\endgroup\@href {#1}{\urlprefix }}%
\providecommand \urlprefix  [0]{URL }%
\providecommand \Eprint [0]{\href }%
\providecommand \doibase [0]{http://dx.doi.org/}%
\providecommand \selectlanguage [0]{\@gobble}%
\providecommand \bibinfo  [0]{\@secondoftwo}%
\providecommand \bibfield  [0]{\@secondoftwo}%
\providecommand \translation [1]{[#1]}%
\providecommand \BibitemOpen [0]{}%
\providecommand \bibitemStop [0]{}%
\providecommand \bibitemNoStop [0]{.\EOS\space}%
\providecommand \EOS [0]{\spacefactor3000\relax}%
\providecommand \BibitemShut  [1]{\csname bibitem#1\endcsname}%
\let\auto@bib@innerbib\@empty
\bibitem [{\citenamefont {Landau}(1937)}]{LS}%
  \BibitemOpen
  \bibfield  {author} {\bibinfo {author} {\bibfnamefont {L.~D.}\ \bibnamefont
  {Landau}},\ }\href@noop {} {\bibfield  {journal} {\bibinfo  {journal} {Zh.
  Eksp. Teor. Fiz.}\ }\textbf {\bibinfo {volume} {7}},\ \bibinfo {pages} {203}
  (\bibinfo {year} {1937})}\BibitemShut {NoStop}%
\bibitem [{\citenamefont {Spitzer}(1962)}]{LS2}%
  \BibitemOpen
  \bibfield  {author} {\bibinfo {author} {\bibfnamefont {L.}~\bibnamefont
  {Spitzer}, \bibfnamefont {Jr.}},\ }\href@noop {} {\emph {\bibinfo {title}
  {Physics of Fully Ionized Gases}}},\ \bibinfo {edition} {2nd}\ ed.\ (\bibinfo
   {publisher} {Interscience},\ \bibinfo {year} {1962})\BibitemShut {NoStop}%
\bibitem [{\citenamefont {Balescu}(1963)}]{LB}%
  \BibitemOpen
  \bibfield  {author} {\bibinfo {author} {\bibfnamefont {R.}~\bibnamefont
  {Balescu}},\ }\href@noop {} {\emph {\bibinfo {title} {Statistical Mechanics
  of Charged Particles}}}\ (\bibinfo  {publisher} {Wiley Interscience},\
  \bibinfo {address} {New York},\ \bibinfo {year} {1963})\BibitemShut {NoStop}%
\bibitem [{\citenamefont {Kremp}\ \emph {et~al.}(2005)\citenamefont {Kremp},
  \citenamefont {Schlanges},\ and\ \citenamefont {Kremp}}]{qLB}%
  \BibitemOpen
  \bibfield  {author} {\bibinfo {author} {\bibfnamefont {D.}~\bibnamefont
  {Kremp}}, \bibinfo {author} {\bibfnamefont {M.}~\bibnamefont {Schlanges}}, \
  and\ \bibinfo {author} {\bibfnamefont {W.-D.}\ \bibnamefont {Kremp}},\
  }\href@noop {} {\emph {\bibinfo {title} {Quantum Statistics of Nonideal
  Plasmas}}}\ (\bibinfo  {publisher} {Springer},\ \bibinfo {address} {Berlin},\
  \bibinfo {year} {2005})\BibitemShut {NoStop}%
\bibitem [{\citenamefont {Gericke}(2005)}]{Gericke05}%
  \BibitemOpen
  \bibfield  {author} {\bibinfo {author} {\bibfnamefont {D.~O.}\ \bibnamefont
  {Gericke}},\ }\href@noop {} {\bibfield  {journal} {\bibinfo  {journal} {J.
  Phys.: Conf. Ser.}\ }\textbf {\bibinfo {volume} {11}},\ \bibinfo {pages}
  {111} (\bibinfo {year} {2005})}\BibitemShut {NoStop}%
\bibitem [{\citenamefont {Daligault}\ and\ \citenamefont
  {Dimonte}(2009)}]{Daligault}%
  \BibitemOpen
  \bibfield  {author} {\bibinfo {author} {\bibfnamefont {J.}~\bibnamefont
  {Daligault}}\ and\ \bibinfo {author} {\bibfnamefont {G.}~\bibnamefont
  {Dimonte}},\ }\href@noop {} {\bibfield  {journal} {\bibinfo  {journal} {Phys.
  Rev. E}\ }\textbf {\bibinfo {volume} {79}},\ \bibinfo {pages} {056403}
  (\bibinfo {year} {2009})}\BibitemShut {NoStop}%
\bibitem [{\citenamefont {Benedict}\ \emph {et~al.}(2012)\citenamefont
  {Benedict}, \citenamefont {Surh}, \citenamefont {Castor}, \citenamefont
  {Khairallah}, \citenamefont {Whitley}, \citenamefont {Richards},
  \citenamefont {Glosli}, \citenamefont {Murillo}, \citenamefont {Scullard},
  \citenamefont {Grabowski}, \citenamefont {Michta},\ and\ \citenamefont
  {Graziani}}]{Benedict}%
  \BibitemOpen
  \bibfield  {author} {\bibinfo {author} {\bibfnamefont {L.~X.}\ \bibnamefont
  {Benedict}}, \bibinfo {author} {\bibfnamefont {M.~P.}\ \bibnamefont {Surh}},
  \bibinfo {author} {\bibfnamefont {J.~I.}\ \bibnamefont {Castor}}, \bibinfo
  {author} {\bibfnamefont {S.~A.}\ \bibnamefont {Khairallah}}, \bibinfo
  {author} {\bibfnamefont {H.~D.}\ \bibnamefont {Whitley}}, \bibinfo {author}
  {\bibfnamefont {D.~F.}\ \bibnamefont {Richards}}, \bibinfo {author}
  {\bibfnamefont {J.~N.}\ \bibnamefont {Glosli}}, \bibinfo {author}
  {\bibfnamefont {M.~S.}\ \bibnamefont {Murillo}}, \bibinfo {author}
  {\bibfnamefont {C.~R.}\ \bibnamefont {Scullard}}, \bibinfo {author}
  {\bibfnamefont {P.~E.}\ \bibnamefont {Grabowski}}, \bibinfo {author}
  {\bibfnamefont {D.}~\bibnamefont {Michta}}, \ and\ \bibinfo {author}
  {\bibfnamefont {F.~R.}\ \bibnamefont {Graziani}},\ }\href@noop {} {\bibfield
  {journal} {\bibinfo  {journal} {Phys. Rev. E}\ }\textbf {\bibinfo {volume}
  {86}},\ \bibinfo {pages} {046406} (\bibinfo {year} {2012})}\BibitemShut
  {NoStop}%
\bibitem [{\citenamefont {Vorberger}\ and\ \citenamefont
  {Gericke}(2009)}]{Vorberger}%
  \BibitemOpen
  \bibfield  {author} {\bibinfo {author} {\bibfnamefont {J.}~\bibnamefont
  {Vorberger}}\ and\ \bibinfo {author} {\bibfnamefont {D.~O.}\ \bibnamefont
  {Gericke}},\ }\href@noop {} {\bibfield  {journal} {\bibinfo  {journal} {Phys.
  Plasmas}\ }\textbf {\bibinfo {volume} {16}},\ \bibinfo {pages} {082702}
  (\bibinfo {year} {2009})}\BibitemShut {NoStop}%
\bibitem [{\citenamefont {Dharma-wardana}\ and\ \citenamefont
  {Perrot}(1998)}]{Dharma-wardana}%
  \BibitemOpen
  \bibfield  {author} {\bibinfo {author} {\bibfnamefont {M.~W.~C.}\
  \bibnamefont {Dharma-wardana}}\ and\ \bibinfo {author} {\bibfnamefont
  {F.}~\bibnamefont {Perrot}},\ }\href@noop {} {\bibfield  {journal} {\bibinfo
  {journal} {Phys. Rev. E}\ }\textbf {\bibinfo {volume} {58}},\ \bibinfo
  {pages} {3705} (\bibinfo {year} {1998})}\BibitemShut {NoStop}%
\bibitem [{\citenamefont {Chapman}\ \emph {et~al.}(2013)\citenamefont
  {Chapman}, \citenamefont {Vorberger},\ and\ \citenamefont
  {Gericke}}]{Chapman}%
  \BibitemOpen
  \bibfield  {author} {\bibinfo {author} {\bibfnamefont {D.~A.}\ \bibnamefont
  {Chapman}}, \bibinfo {author} {\bibfnamefont {J.}~\bibnamefont {Vorberger}},
  \ and\ \bibinfo {author} {\bibfnamefont {D.~O.}\ \bibnamefont {Gericke}},\
  }\href@noop {} {\bibfield  {journal} {\bibinfo  {journal} {Phys. Rev. E}\
  }\textbf {\bibinfo {volume} {88}},\ \bibinfo {pages} {013102} (\bibinfo
  {year} {2013})}\BibitemShut {NoStop}%
\bibitem [{\citenamefont {Atzeni}\ and\ \citenamefont
  {Meyer{-}ter{-}Vehn}(2004)}]{ICF}%
  \BibitemOpen
  \bibfield  {author} {\bibinfo {author} {\bibfnamefont {S.}~\bibnamefont
  {Atzeni}}\ and\ \bibinfo {author} {\bibfnamefont {J.}~\bibnamefont
  {Meyer{-}ter{-}Vehn}},\ }\href@noop {} {\emph {\bibinfo {title} {The Physics
  of Inertial Fusion}}}\ (\bibinfo  {publisher} {Clarendon},\ \bibinfo
  {address} {Oxford},\ \bibinfo {year} {2004})\BibitemShut {NoStop}%
\bibitem [{\citenamefont {Benedict}\ \emph {et~al.}(2017)\citenamefont
  {Benedict}, \citenamefont {Surh}, \citenamefont {Stanton}, \citenamefont
  {Scullard}, \citenamefont {Correa}, \citenamefont {Castor}, \citenamefont
  {Graziani}, \citenamefont {Collins}, \citenamefont {\u{C}ert\'{\i}k},
  \citenamefont {Kress},\ and\ \citenamefont {Murillo}}]{Benedict17}%
  \BibitemOpen
  \bibfield  {author} {\bibinfo {author} {\bibfnamefont {L.~X.}\ \bibnamefont
  {Benedict}}, \bibinfo {author} {\bibfnamefont {M.~P.}\ \bibnamefont {Surh}},
  \bibinfo {author} {\bibfnamefont {L.~G.}\ \bibnamefont {Stanton}}, \bibinfo
  {author} {\bibfnamefont {C.~R.}\ \bibnamefont {Scullard}}, \bibinfo {author}
  {\bibfnamefont {A.~A.}\ \bibnamefont {Correa}}, \bibinfo {author}
  {\bibfnamefont {J.~I.}\ \bibnamefont {Castor}}, \bibinfo {author}
  {\bibfnamefont {F.~R.}\ \bibnamefont {Graziani}}, \bibinfo {author}
  {\bibfnamefont {L.~A.}\ \bibnamefont {Collins}}, \bibinfo {author}
  {\bibfnamefont {O.}~\bibnamefont {\u{C}ert\'{\i}k}}, \bibinfo {author}
  {\bibfnamefont {J.~D.}\ \bibnamefont {Kress}}, \ and\ \bibinfo {author}
  {\bibfnamefont {M.~S.}\ \bibnamefont {Murillo}},\ }\href@noop {} {\bibfield
  {journal} {\bibinfo  {journal} {Phys. Rev. E}\ }\textbf {\bibinfo {volume}
  {95}},\ \bibinfo {pages} {043202} (\bibinfo {year} {2017})}\BibitemShut
  {NoStop}%
\bibitem [{\citenamefont {Garbett}\ and\ \citenamefont
  {Chapman}(2016)}]{Garbett}%
  \BibitemOpen
  \bibfield  {author} {\bibinfo {author} {\bibfnamefont {W.~J.}\ \bibnamefont
  {Garbett}}\ and\ \bibinfo {author} {\bibfnamefont {D.~A.}\ \bibnamefont
  {Chapman}},\ }\href@noop {} {\bibfield  {journal} {\bibinfo  {journal} {J.
  Phys.: Conf. Ser.}\ }\textbf {\bibinfo {volume} {688}},\ \bibinfo {pages}
  {012019} (\bibinfo {year} {2016})}\BibitemShut {NoStop}%
\bibitem [{\citenamefont {Ichimaru}\ \emph {et~al.}(1985)\citenamefont
  {Ichimaru}, \citenamefont {Mitake}, \citenamefont {Tanaka},\ and\
  \citenamefont {Yan}}]{Ichimaru}%
  \BibitemOpen
  \bibfield  {author} {\bibinfo {author} {\bibfnamefont {S.}~\bibnamefont
  {Ichimaru}}, \bibinfo {author} {\bibfnamefont {S.}~\bibnamefont {Mitake}},
  \bibinfo {author} {\bibfnamefont {S.}~\bibnamefont {Tanaka}}, \ and\ \bibinfo
  {author} {\bibfnamefont {X.-Z.}\ \bibnamefont {Yan}},\ }\href@noop {}
  {\bibfield  {journal} {\bibinfo  {journal} {Phys. Rev. A}\ }\textbf {\bibinfo
  {volume} {32}},\ \bibinfo {pages} {1768} (\bibinfo {year}
  {1985})}\BibitemShut {NoStop}%
\bibitem [{\citenamefont {Dawson}(1897)}]{Dawson}%
  \BibitemOpen
  \bibfield  {author} {\bibinfo {author} {\bibfnamefont {H.~G.}\ \bibnamefont
  {Dawson}},\ }\href@noop {} {\bibfield  {journal} {\bibinfo  {journal} {Proc.
  London Math. Soc.}\ }\textbf {\bibinfo {volume} {s1-29}} (\bibinfo {year}
  {1897})}\BibitemShut {NoStop}%
\bibitem [{\citenamefont {Stanton}\ and\ \citenamefont
  {Murillo}(2016)}]{Stanton}%
  \BibitemOpen
  \bibfield  {author} {\bibinfo {author} {\bibfnamefont {L.~G.}\ \bibnamefont
  {Stanton}}\ and\ \bibinfo {author} {\bibfnamefont {M.~S.}\ \bibnamefont
  {Murillo}},\ }\href@noop {} {\bibfield  {journal} {\bibinfo  {journal} {Phys.
  Rev. E}\ }\textbf {\bibinfo {volume} {93}},\ \bibinfo {pages} {043203}
  (\bibinfo {year} {2016})}\BibitemShut {NoStop}%
\bibitem [{\citenamefont {Dandrea}\ \emph {et~al.}()\citenamefont {Dandrea},
  \citenamefont {Ashcroft},\ and\ \citenamefont {Carlsson}}]{Dandrea}%
  \BibitemOpen
  \bibfield  {author} {\bibinfo {author} {\bibfnamefont {R.~G.}\ \bibnamefont
  {Dandrea}}, \bibinfo {author} {\bibfnamefont {N.~W.}\ \bibnamefont
  {Ashcroft}}, \ and\ \bibinfo {author} {\bibfnamefont {A.~E.}\ \bibnamefont
  {Carlsson}},\ }\href@noop {} {\bibinfo  {journal} {Phys. Rev. B}\
  }\BibitemShut {NoStop}%
\bibitem [{\citenamefont {Oberman}\ \emph {et~al.}(1963)\citenamefont
  {Oberman}, \citenamefont {Ron},\ and\ \citenamefont {Dawson}}]{Oberman}%
  \BibitemOpen
\bibfield  {journal} {  }\bibfield  {author} {\bibinfo {author} {\bibfnamefont
  {C.}~\bibnamefont {Oberman}}, \bibinfo {author} {\bibfnamefont
  {A.}~\bibnamefont {Ron}}, \ and\ \bibinfo {author} {\bibfnamefont
  {J.}~\bibnamefont {Dawson}},\ }\href@noop {} {\bibfield  {journal} {\bibinfo
  {journal} {Phys. Fluids}\ }\textbf {\bibinfo {volume} {5}},\ \bibinfo {pages}
  {3705} (\bibinfo {year} {1963})}\BibitemShut {NoStop}%
\bibitem [{\citenamefont {Williams}\ and\ \citenamefont
  {DeWitt}(1969)}]{Williams}%
  \BibitemOpen
  \bibfield  {author} {\bibinfo {author} {\bibfnamefont {R.~H.}\ \bibnamefont
  {Williams}}\ and\ \bibinfo {author} {\bibfnamefont {H.~E.}\ \bibnamefont
  {DeWitt}},\ }\href@noop {} {\bibfield  {journal} {\bibinfo  {journal} {Phys.
  Fluids}\ }\textbf {\bibinfo {volume} {12}},\ \bibinfo {pages} {2326}
  (\bibinfo {year} {1969})}\BibitemShut {NoStop}%
\bibitem [{\citenamefont {Whitley}\ \emph {et~al.}(2015)\citenamefont
  {Whitley}, \citenamefont {Scullard}, \citenamefont {Benedict}, \citenamefont
  {Castor}, \citenamefont {Randles}, \citenamefont {Glosli}, \citenamefont
  {Richards}, \citenamefont {Desjarlais},\ and\ \citenamefont
  {Graziani}}]{Whitley}%
  \BibitemOpen
  \bibfield  {author} {\bibinfo {author} {\bibfnamefont {H.~D.}\ \bibnamefont
  {Whitley}}, \bibinfo {author} {\bibfnamefont {C.~R.}\ \bibnamefont
  {Scullard}}, \bibinfo {author} {\bibfnamefont {L.~X.}\ \bibnamefont
  {Benedict}}, \bibinfo {author} {\bibfnamefont {J.~I.}\ \bibnamefont
  {Castor}}, \bibinfo {author} {\bibfnamefont {A.}~\bibnamefont {Randles}},
  \bibinfo {author} {\bibfnamefont {J.~N.}\ \bibnamefont {Glosli}}, \bibinfo
  {author} {\bibfnamefont {D.~F.}\ \bibnamefont {Richards}}, \bibinfo {author}
  {\bibfnamefont {M.~P.}\ \bibnamefont {Desjarlais}}, \ and\ \bibinfo {author}
  {\bibfnamefont {F.~R.}\ \bibnamefont {Graziani}},\ }\href@noop {} {\bibfield
  {journal} {\bibinfo  {journal} {Contributions to Plasma Physics}\ }\textbf
  {\bibinfo {volume} {55}},\ \bibinfo {pages} {192} (\bibinfo {year}
  {2015})}\BibitemShut {NoStop}%
\bibitem [{\citenamefont {Managan}(2015)}]{Managan}%
  \BibitemOpen
  \bibfield  {author} {\bibinfo {author} {\bibfnamefont {R.~A.}\ \bibnamefont
  {Managan}},\ }\href {https://e-reports-ext.llnl.gov/pdf/787581.pdf}
  {\bibfield  {journal} {\bibinfo  {journal} {NECDC: 18 Biennial Nuclear
  Explosives Code Development Conference, United States}\ } (\bibinfo {year}
  {2015})}\BibitemShut {NoStop}%
\bibitem [{\citenamefont {Brysk}(1974)}]{Brysk}%
  \BibitemOpen
  \bibfield  {author} {\bibinfo {author} {\bibfnamefont {H.}~\bibnamefont
  {Brysk}},\ }\href@noop {} {\bibfield  {journal} {\bibinfo  {journal} {Plasma
  Phys.}\ }\textbf {\bibinfo {volume} {16}},\ \bibinfo {pages} {927} (\bibinfo
  {year} {1974})}\BibitemShut {NoStop}%
\end{thebibliography}%

\appendix

\section{Fitting functions for Fermi integrals} \label{sec:Fermi_functions}
Here, we give the fits for computing the chemical potential and the Fermi integral $I_{-1/2}(-\mu)$.

For the chemical potential, we use the fit given by Managan \cite{Managan},
\begin{equation}
 \mu = - \ln(e^{R_3(\xi)} -1)
\end{equation}
with
\begin{equation}
 R_3(\xi) \equiv \frac{\frac{4}{3 \sqrt{\pi}}+\sum_{i=1}^3 a_i \xi^i}{1+\sum_{i=1}^4 b_i \xi^i} \label{eq:R3}
\end{equation}
and
\begin{equation}
 \xi \equiv \Theta^{-\frac{1}{2}}
\end{equation}
where $\Theta$ is the standard degeneracy parameter, given in (\ref{eq:Theta}), and the coefficients $a_i$ and $b_i$ are
\begin{eqnarray}
 a_1 &=& 0.19972  \cr
 a_2 &=& 0.17258  \cr
 a_3 &=& 0.145    \cr
 b_1 &=& 0.25829  \cr
 b_2 &=& 0.28756  \cr
 b_3 &=& 0.16842  \cr
 b_4 &=& 0.145 \ .  \nonumber
\end{eqnarray}
Note that if $R_3$ reaches a certain size, say $R_3>5$, then one can just set $\mu=-R_3$.

For the Fermi integral $I_{-\frac{1}{2}}(-\mu)$, we use the formula of Dandrea, Ashcroft and Carlsson \cite{Dandrea}, good for all values of $\Theta$,
\begin{widetext}
\begin{equation}
 I_{-\frac{1}{2}}(-\mu)=\frac{2}{\sqrt{\pi \Theta}} \frac{1+c_1 \Theta^2+c_2 \Theta^4+c_3 \Theta^6}{1+(c_1+\pi^2/12)\Theta^2+c_4 \Theta^4+(c_3/\sqrt{2 \pi})\Theta^{11/2}+(3 c_3/2)\Theta^7}
\end{equation}
\end{widetext}
where
\begin{eqnarray}
 c_1&=&41.775\\
 c_2&=&27.390\\
 c_3&=&4287.2\\
 c_4&=&50.605 \ .
\end{eqnarray}

\section{Response functions without quantum diffraction} \label{sec:response_functions}
The element that makes the dielectric function difficult to deal with is quantum diffraction. Without it, there is a separation of the variables $x$ and $y$, as in equation (\ref{eq:dielectric}), even when we include the effects of degeneracy. The quantum free-particle response function is given by (\ref{eq:chi}) and we neglect diffraction by taking the limit $\hbar \rightarrow 0$. Thus, we use for the electrons
\begin{equation}
 \chi_e(k,\omega) = -\frac{1}{m_e} \lim_{\eta \rightarrow 0^+} \int d^3 \bfv \frac{\bfk \cdot \nabla_{\bfv} f_e(\bfv)}{\omega - \bfv \cdot \bfk +i \eta} . \label{eq:chic}
\end{equation}
where $f(\bfv)$ is the Fermi-Dirac distribution, equation (\ref{eq:FD}), in which we do {\it not} take $\hbar \rightarrow 0$. We then have
\begin{eqnarray}
 & &\chi_e(k,\omega) = \frac{4 \pi m_e^3 \beta_e e^\mu}{h^3} \cr
 &\times& \int_{-\infty}^{\infty} \int_0^\infty \frac{k v_z e^{\beta_e m_e v^2/2} v_\perp d v_\perp d v_z}{(\omega-k v_z +i \eta)(e^{\mu+\beta_e m_e v^2/2}+1)^2} 
\end{eqnarray}
which, by noting that $v^2=v_\perp^2+v_z^2$, can be written
\begin{equation}
 \chi_e(k,\omega) = \frac{4 \pi m_e^3 \beta_e e^\mu}{h^3} \int_{-\infty}^{\infty} \frac{k v_z e^{\beta_e m_e v_z^2/2} I_\perp(v_z) dv_z}{(\omega-k v_z +i \eta)} 
\end{equation}
where
\begin{eqnarray}
 I_\perp(v_z) &\equiv& \int_0^\infty \frac{e^{\beta_e m_e v_\perp^2/2} v_\perp d v_\perp}{(e^{\mu+\beta_e m_e v_\perp^2/2+\beta_e m_e v_z^2/2}+1)^2} \\
 &=& \frac{1}{\beta_e m_e} \int_0^\infty \frac{e^x dx}{(e^{\mu+x+\beta_e m_e v_z^2/2}+1)^2} \\
 &=& \frac{1}{\beta_e m_e} \frac{e^{-\mu}e^{-\beta_e m_e v_z^2/2}}{e^{\mu + \beta_e m_e v_z^2/2}+1} .
\end{eqnarray}
This leaves
\begin{equation}
 \chi_e(x) = \frac{4 \pi m_e^2}{h^3} \sqrt{\frac{2}{\beta_e m_e}} \int_{-\infty}^{\infty} \frac{z dz}{(\sqrt{\alpha}x-z+i \eta)(e^\mu e^{z^2}+1)} 
\end{equation}
from which the Sokhotski-Plemelj theorem immediately gives
\begin{equation}
 \mathrm{Im} \chi_e(x)= -\frac{4 \pi^2 m_e^2}{h^3} \sqrt{\frac{2}{\beta_e m_e}} \sqrt{\alpha}x \frac{1}{1+e^\mu e^{\alpha x^2}} .
\end{equation}
To get the real part, we use the usual trick
\begin{equation}
 \frac{1}{(\sqrt{\alpha}x-z+i \eta)}=-i \int_0^\infty e^{i (\sqrt{\alpha}x-z)t}e^{-\eta t}dt
\end{equation}
to get
\begin{equation}
 \chi_e(x) = -i \frac{4 \pi m_e^2}{h^3} \sqrt{\frac{2}{\beta_e m_e}} \int_0^\infty \int_{-\infty}^{\infty} \frac{z e^{i (\sqrt{\alpha}x-z)t}e^{-\eta t}dz dt}{e^\mu e^{z^2}+1} \ . 
\end{equation}
The easiest way to proceed appears to be to use the expansion
\begin{equation}
 \frac{1}{e^\mu e^{z^2}+1}=\sum_{n=0}^\infty (-1)^n e^{-(n+1)\mu}e^{-(n+1)z^2}
\end{equation}
which is only valid when $\mu>0$. At the end we will get an expression that is valid for the whole range of $\mu$ and we will claim it is correct by analytic extension. Using the expansion we get
\begin{eqnarray}
& &\chi_e(x) = -i \frac{4 \pi m_e^2}{h^3} \sqrt{\frac{2}{\beta_e m_e}} \sum_{n=0}^\infty (-1)^n e^{-(n+1)\mu} \cr 
&\times&\int_0^\infty \int_{-\infty}^{\infty} z e^{-(n+1)z^2} e^{i (\sqrt{\alpha}x-z)t} e^{-\eta t}dz dt \ ,
\end{eqnarray}
and the $z$-integral is
\begin{eqnarray}
& &\int_{\infty}^\infty e^{-(n+1)z^2}z e^{-i z t}dz= \cr
 &-&\frac{i \sqrt{\pi}}{2 (n+1)^{3/2}} t \exp \left(-\frac{t^2}{4(n+1)} \right)
\end{eqnarray}
so that
\begin{eqnarray}
& &\chi_e(x) = - \frac{2 \pi^{3/2} m_e^2}{h^3} \sqrt{\frac{2}{\beta_e m_e}} \sum_{n=0}^\infty (-1)^n \frac{e^{-(n+1)\mu}}{(n+1)^{3/2}} \cr
&\times& \int_0^\infty  t \exp \left(-\frac{t^2}{4(n+1)} \right) e^{i x t} dt \,
\end{eqnarray}
where we have set $\eta=0$ because it is no longer needed. Now,
\begin{eqnarray}
& &\int_0^\infty  t \exp \left(-\frac{t^2}{4(n+1)} \right) e^{i \sqrt{\alpha} x t} dt = \cr
& &2(1+n) \left[1-2 \sqrt{1+n} \sqrt{\alpha} x F(\sqrt{n+1}\sqrt{\alpha} x) \right. \cr
 &+& \left. i \sqrt{n+1} \sqrt{\pi} e^{-(n+1) \alpha x^2} \right]
\end{eqnarray}
where $F(x)$ is the Dawson function, equation (\ref{eq:Dawson}), so
\begin{eqnarray}
& &\chi_e(x) = - \frac{4 \pi^{3/2} m_e^2}{h^3} \sqrt{\frac{2}{\beta_e m_e}} \sum_{n=0}^\infty (-1)^n e^{-(n+1)\mu} \cr
&\times& \bigg[\frac{1}{(n+1)^{1/2}} - 2 \sqrt{\alpha} x F(\sqrt{n+1}\sqrt{\alpha} x) \cr
& & \ \ \ \ \ \ \ \ \ \ \ \ \ \ \ \ \ \ \ \ \ \ \ \ \ \ +i \sqrt{\pi} \sqrt{\alpha} x e^{-(n+1)\alpha x^2} \bigg]. \label{eq:chi_e_sum}
\end{eqnarray}
Under the assumption $\mu>0$, these sums can be done exactly. From the definition of the polylogarithm $\mathrm{Li}_s(x)$, we have
\begin{eqnarray}
& &\sum_{n=0}^\infty (-1)^n \frac{e^{-(n+1) \mu}}{\sqrt{1+n}} =-\mathrm{Li}_{1/2}(-e^{-\mu}) \cr
 &=& \frac{1}{\sqrt{\pi}} \int_0^{\infty} \frac{t^{-1/2}}{e^{t+\mu}+1} dt = I_{-1/2}(-\mu)\ . \label{eq:Iminhalf}
\end{eqnarray}
The definition of $\mathrm{Li}_s(x)$ is only valid when $|x|<1$, and the Fermi integral provides the analytic extension to $x<-1$, which is what we need when $\mu<0$. Next we define
\begin{eqnarray}
 & &\sum_{n=0}^{\infty} e^{-n \mu} (-1)^n F(\sqrt{1+n} \sqrt{\alpha}x) \equiv F(\sqrt{\alpha} x;\mu) \label{eq:genDawson}
\end{eqnarray}
where $F(x;\mu)$ is a generalization of the Dawson function with the property $\lim_{\mu \rightarrow \infty} F(x;\mu)=F(x)$. We could now use (\ref{eq:Dawson}) in (\ref{eq:genDawson}) to find an integral form for $F(x;\mu)$. However, as we show in the main text, we can set $\alpha=0$ in the real part of the dielectric function, even in the coupled mode regime, so we will not bother with this. We can now write the real part of the response function,
\begin{eqnarray}
 \mathrm{Re}\chi_e(x) &=& -\frac{4 \pi^{3/2} m_e^2 e^{-\mu}}{h^3} \sqrt{\frac{2}{\beta_e m_e}} \cr
 &\times&\left[I_{-\frac{1}{2}}(-\mu)e^\mu-2 \sqrt{\alpha}x F(\sqrt{\alpha} x;\mu) \right]. \label{eq:Rechi_e_nodiff}
\end{eqnarray}
In the classical limit, $\mu$ blows up as
\begin{equation}
 e^{-\mu}=\frac{n_e h^3}{2 \pi^{3/2} m_e^3}\left(\frac{\beta_e m_e}{2}\right)^{3/2}
\end{equation}
and $I_{-1/2}(-\mu)e^\mu \rightarrow 1$ so we have
\begin{equation}
 \mathrm{Re}\chi_e(x) = - n_e \beta_e \left[1-2 \sqrt{\alpha} x F(\sqrt{\alpha}x)\right] .
\end{equation}
If $\alpha=0$, this is simply $-n_e \beta_e$, which strongly suggests that we write
\begin{equation}
 \mathrm{Re}\chi_e(x) = - n_e \beta_{\mathrm{eff}} \label{eq:Rechi_e_static}
\end{equation}
where the effective temperature $\beta_{\mathrm{eff}}$ is easily read from (\ref{eq:Rechi_e_nodiff}) and is discussed in the main text.

\section{Brysk formula} \label{sec:Brysk}
The Brysk correction \cite{Brysk} to the Landau-Spitzer formula was derived mainly from collisional arguments. Here, we show that it is possible to arrive at the identical formula from the Lenard-Balescu integral (\ref{eq:dTdti}) by neglecting quantum diffraction everywhere and setting the dielectric function to $1$. Mathematically, quantum diffraction is neglected in equation (\ref{eq:integral_sinh}) simply by setting $y=0$ everywhere, except in $dy/y$. The reason for this is that the dimensionless wavenumber, $y$, contains the factor of $\hbar$ arising from quantum diffraction, but it cancels out of $dy/y$. Then  setting
\begin{equation}
 \int_{y_{\min}}^{y_{\max}}\frac{dy}{y}=\ln \Lambda,
\end{equation}
equation (\ref{eq:integral_sinh}) becomes
\begin{equation}
 \frac{d T_i}{dt}=\frac{8}{3}\frac{Z^2 e^4 m_e^2}{\pi \hbar^3 m_i} \frac{1}{1+e^\mu} \ln \Lambda(T_e-T_i) \label{eq:Brysk}
\end{equation}
which is identical with equation (35) of \cite{Brysk}, where Brysk's $A$ is our $e^\mu$.

\section{Large $z$ expansion of the dielectric function} \label{sec:Laurent}
Here, we derive the coefficients $a_{2n}(y)$ appearing in equation (\ref{eq:epsexp}). First, we need to find the expansion for the ion response function. 
Using the Maxwell distribution and our variables $x$ and $y$, we can write it in the following form
\begin{equation}
 \chi_i(x)= \frac{\beta_i n_i}{\sqrt{\pi}}\int_{-\infty}^\infty \frac{u e^{-u^2} du}{x-u}
\end{equation}
and in the complex plane,
\begin{equation}
 \chi_i(z)= \frac{\beta_i n_i}{\sqrt{\pi}}\frac{1}{z}\int_{-\infty}^\infty \frac{u e^{-u^2} du}{1-u/z} .
\end{equation}
Now we use the expansion
\begin{equation}
 \frac{1}{1-u/z}=\sum_{n=0}^\infty \left(\frac{u}{z}\right)^n .
\end{equation}
The response function becomes
\begin{equation}
 \chi_i(z)=\frac{\beta_i n_i}{\sqrt{\pi}}\sum_{n=0}^{\infty}\frac{1}{z^{n+1}} \int_{-\infty}^\infty u^{n+1} e^{-u^2} du
\end{equation}
where
\begin{equation}
 \int_{-\infty}^\infty u^{n+1} e^{-u^2} du = \frac{1}{2} n \Gamma\left(\frac{n}{2}\right)
\end{equation}
for $n$ odd and is zero otherwise, where $\Gamma(n)$ is the gamma function. Putting these together, we get
\begin{equation}
 \chi_i(z)=\frac{\beta_i n_i}{2 \sqrt{\pi}}\sum_{n=0}^{\infty}\frac{(2n+1)\Gamma(n+1/2)}{z^{2n+2}} . \label{eq:chi_iz}
\end{equation}
The dielectric function is given by (\ref{eq:dielectric}), where $w(x)$ is essentially the sum of the electron and ion response functions (as in (\ref{eq:eps_definition}) but in dimensionless variables). Inserting the expression (\ref{eq:chi_iz}) for the ions and setting $\alpha=0$ for the electrons we find
\begin{equation}
 \frac{\eta^2 \gammaeff}{y^2 Z \epsilon(z,y)}=\frac{1}{\zeta y^2+1-\sum_{n=0}^\infty A_{2n+1}/z^{2n+2}} \label{eq:inveps}
\end{equation}
where
\begin{equation}
 A_n \equiv \frac{Z}{2 \sqrt{\pi} \gammaeff} n \Gamma\left(\frac{n}{2}\right) ,
\end{equation}
and
\begin{equation}
 \zeta \equiv \frac{Z}{\gammaeff \eta^2}.
\end{equation}
We can now use equation (\ref{eq:inveps}) to calculate the coefficients $a_{2n}(\tau)$ of the asymptotic expansion (\ref{eq:epsexp}). This is done by an expansion in the small quantity
\begin{equation}
 \frac{1}{\zeta y^2+1}\sum_{n=0}^\infty \frac{A_{2n+1}}{z^{2n+2}} .
\end{equation}
The result is
\begin{equation}
 a_{2n}(y)=\frac{1}{2^n} \frac{1}{\zeta y^2+1} P_{n}\left(\frac{Z}{\gammaeff}\frac{1}{\zeta y^2+1} \right)
\end{equation}
where $P_n(w)$ are a set of polynomials. The first few of these are given by
\begin{eqnarray}
 P_0(w)&=&1 \cr
 P_1(w)&=&w \cr
 P_2(w)&=&w^2+3w \cr
 P_3(w)&=&w^3+6w^2+15 w \cr
 P_4(w)&=&w^4+9w^3+39 w^2+105w \cr
 P_5(w)&=&w^5+12w^4+72 w^3+300w^2+945w .
\end{eqnarray}
\section{Series expansions of special functions} \label{sec:special_func}
Here, we report the series expansions of the special functions used throughout the paper. We will derive only (\ref{eq:fser}) as the procedure is the same for the others.

Making the substitution $y=t+x$, (\ref{eq:ftilde}) becomes
\begin{equation}
 \tilde{f}(x;\mu)=e^x [\tilde{f}_2(x;\mueff)-x \tilde{f}_3(x;\mueff)] \label{eq:fdecomp}
\end{equation}
where
\begin{eqnarray}
 \tilde{f}_2(x;\mu)&\equiv&\int_x^\infty \frac{e^{-y}}{y(1+e^{-\mu-y})}dy \\
 \tilde{f}_3(x;\mu)&\equiv&\int_x^\infty \frac{e^{-y}}{y^2(1+e^{-\mu-y})}dy , \label{eq:f3tilde}
\end{eqnarray}
and $\mueff \equiv \mu-x$. One might as well just use this effective chemical potential rather than expanding the exponential in the denominator. We will focus on $\tilde{f}_3(x;\mu)$ because $\tilde{f}_2(x)$ is handled the same way. To get most of the terms, we find the power series of the derivative of (\ref{eq:f3tilde}),
\begin{equation}
 \frac{d \tilde{f}_3(x;\mu)}{dx}=-\frac{e^{\mu}}{B(\mu)}\frac{1}{x^2}+\frac{e^{2\mu}}{[B(\mu)]^2}\frac{1}{x}-\frac{e^{2\mu}(e^{\mu}-1)}{2 [B(\mu)]^3}+ ...
\end{equation}
where $B(\mu)$ is defined in (\ref{eq:B}), and integrate this term by term. This does not, however, determine the constant (order unity) term. To get this, we integrate (\ref{eq:f3tilde}) by parts to obtain
\begin{equation}
 \tilde{f}_3(x;\mu)=\frac{1}{x}\frac{e^{-x}}{1+e^{-\mu-x}}-\int_x^\infty \frac{e^{-t}}{t(1+e^{-\mu-t})^2}dt \ .
\end{equation}
Integrating the second term by parts gives
\begin{eqnarray}
 \tilde{f}_3(x;\mu)&=&\frac{1}{x}\frac{e^{-x}}{1+e^{-\mu-x}}+\ln x \frac{e^{-x}}{(1+e^{-\mu-x})^2} \cr
 &-&\int_x^\infty \ln t \frac{(e^{-t}-e^{-\mu-2t})}{(1+e^{-\mu-t})^3}dt \ ,
\end{eqnarray}
and now the integral in this expression is convergent as $x \rightarrow 0$. We can simply take this limit to obtain
\begin{equation}
 \tilde{f}_3(x;\mu) \rightarrow U_2(\mu)-\frac{e^{2 \mu}}{[B(\mu)]^2}+\ln x \frac{e^{2 \mu}}{[B(\mu)]^2}
\end{equation}
as $x \rightarrow 0$, where $U_2(\mu)$ is defined by (\ref{eq:C3}). We have thus obtained the series of $\tilde{f}_3$. Doing the same procedure on $\tilde{f}_2$ and plugging these results into (\ref{eq:fdecomp}), we obtain (\ref{eq:fser}).

\section{Series expansions for $\overline{\Gamma}_n$ integrals} \label{sec:Gammaseries}
Below we present the series expansions for the integrals $\overline{\Gamma}^{(j)}_n(\gammaeff/Z,0)$ needed for the coupled mode correction factor (\ref{eq:DeltaR}). These are derived by means similar to those used to compute the dielectric function integrals exactly. It is somewhat more complicated, but we will not go into the details. The series below are generally good for $\gammaeff/Z<0.1$ and outside this range one probably need not worry about coupled modes. The following should be sufficient for nearly all applications,
\begin{eqnarray}
 \overline{\Gamma}_0^{(0)}&=&-\sqrt{\pi}c_1^{(0)}(\gammaeff/Z)+0.53485881+0.79488185 \frac{\gammaeff}{Z} \cr
 &+& 0.79488185 \left(\frac{\gammaeff}{Z} \right)^2 + 4.3835505 \left(\frac{\gammaeff}{Z} \right)^3 \\
 \overline{\Gamma}_1^{(0)}&=&-\sqrt{\pi}c_2^{(0)}(\gammaeff/Z)+1.2910742+3.0889424 \frac{\gammaeff}{Z} \cr
 &+& 6.1778849 \left(\frac{\gammaeff}{Z} \right)^2 + 32.058147 \left(\frac{\gammaeff}{Z} \right)^3 \\
 \overline{\Gamma}_0^{(1)}&=&-\sqrt{\pi}c_1^{(1)}(\gammaeff/Z)+0.51708588-0.10157305 \frac{\gammaeff}{Z} \cr
 &+& 1.0154030 \left(\frac{\gammaeff}{Z} \right)^2 + 0.67693539 \left(\frac{\gammaeff}{Z} \right)^3 \\
 \overline{\Gamma}_1^{(1)}&=&-\sqrt{\pi}c_2^{(1)}(\gammaeff/Z)+0.32092230+0.60469118 \frac{\gammaeff}{Z} \cr
 &+& 3.7222584 \left(\frac{\gammaeff}{Z} \right)^2 + 4.9630113 \left(\frac{\gammaeff}{Z} \right)^3 \\
  \overline{\Gamma}_0^{(2)}&=&-\sqrt{\pi}c_1^{(2)}(\gammaeff/Z)-0.036391827+1.5543671 \frac{\gammaeff}{Z} \cr
 &-& 1.2959555 \left(\frac{\gammaeff}{Z} \right)^2 + 1.0930122 \left(\frac{\gammaeff}{Z} \right)^3 \\
 \overline{\Gamma}_1^{(2)}&=&-\sqrt{\pi}c_2^{(2)}(\gammaeff/Z)-0.28938771 + 1.5376747 \frac{\gammaeff}{Z} \cr
 &-& 0.92477069 \left(\frac{\gammaeff}{Z} \right)^2 + 3.9070984 \left(\frac{\gammaeff}{Z} \right)^3 \\
   \overline{\Gamma}_0^{(3)}&=&-\sqrt{\pi}c_1^{(3)}(\gammaeff/Z)+0.052145343-0.19030541 \frac{\gammaeff}{Z} \cr
 &+& 3.1112345 \left(\frac{\gammaeff}{Z} \right)^2 -2.8365990 \left(\frac{\gammaeff}{Z} \right)^3 \\
 \overline{\Gamma}_1^{(3)}&=&-\sqrt{\pi}c_2^{(3)}(\gammaeff/Z)+0.062897190 -1.1956660 \frac{\gammaeff}{Z} \cr
 &+& 3.9093434 \left(\frac{\gammaeff}{Z} \right)^2 -2.9563191 \left(\frac{\gammaeff}{Z} \right)^3 ,
\end{eqnarray}
where
\begin{eqnarray}
c_1^{(0)}&=&-\frac{1}{2}\left(\log \frac{\gammaeff}{Z}+1 \right) \\
c_1^{(1)}&=&-\frac{1}{2} \frac{\gammaeff}{Z}\left(2\log \frac{\gammaeff}{Z}+1 \right) \\
c_1^{(2)}&=&-\frac{1}{2}\left(\frac{\gammaeff}{Z}\right)^2\left(3\log \frac{\gammaeff}{Z}+1 \right) \\
c_1^{(3)}&=&-\frac{1}{2}\left(\frac{\gammaeff}{Z}\right)^3\left(4\log \frac{\gammaeff}{Z}+1 \right) \\
c_2^{(0)}&=&-\frac{3}{4}+\frac{Z}{8 \gammaeff}-\frac{3}{4}\log \frac{\gammaeff}{Z} \\
c_2^{(1)}&=&\frac{3}{8}-\frac{3 \gammaeff}{4Z}+ \frac{1}{4}\left(1-6 \frac{\gammaeff}{Z} \right) \log \frac{\gammaeff}{Z} \\
c_2^{(2)}&=&-\frac{3}{4}\frac{\gammaeff^2}{Z^2}+\frac{5 \gammaeff}{8Z} \cr
& & \ \ \ \ \ \ \ \ + \frac{3 \gammaeff}{4 Z}\left(1-3 \frac{\gammaeff}{Z} \right) \log \frac{\gammaeff}{Z} \\
c_2^{(3)}&=&-\frac{3}{4}\frac{\gammaeff^3}{Z^3}+\frac{7 \gammaeff^2}{8Z^2} \cr
& & \ \ \ \ \ \ \ \ + \frac{3 \gammaeff^2}{2 Z^2}\left(1-2 \frac{\gammaeff}{Z} \right) \log \frac{\gammaeff}{Z}.
\end{eqnarray}
\end{document}